\documentclass[aps,amsmath,amssymb,twocolumn,10pt,superscriptaddress,pra]{revtex4-1}
\usepackage[T1]{fontenc}
\usepackage{bbold}
\usepackage{graphicx}  
\usepackage[colorlinks=true,linkcolor=blue,citecolor=red,urlcolor=magenta]{hyperref}
\begin{document} 

\title{Impurity immersed in a two-component few-fermion mixture \\in a one-dimensional harmonic trap} 

\author{Marek Teske}
\affiliation{\mbox{Institute of Physics, Polish Academy of Sciences, Aleja Lotnikow 32/46, PL-02668 Warsaw, Poland}}
\author{Tomasz Sowi\'nski}
\affiliation{\mbox{Institute of Physics, Polish Academy of Sciences, Aleja Lotnikow 32/46, PL-02668 Warsaw, Poland}}
 
\begin{abstract}
We investigate a one-dimensional three-component few-fermion mixture confined in a parabolic external trap, where one component contains a single particle acting as an impurity. Focusing on the many-body ground state, we analyze how the interactions between the impurity and the other components influence the system's structure. For fixed interaction strengths within the mixture, we identify a critical interaction strength with the impurity for which the system undergoes a structural transition characterized by a substantial change in its spatial features. We explore this transition from the point of view of correlations and ground-state susceptibility. We remarkably find that this transition exhibits unique universality features not previously observed in other systems, highlighting novel many-body properties existing in multi-component fermionic mixtures.
\end{abstract}
\maketitle
\section{Introduction}
One of the most amazing puzzles of nature is the collective manifestation of quantum phenomena in the macroworld. Spectacular examples of such behaviour include the superfluidity~\cite{1938KapitzaNature,1938AllenNature}, the superconductivity~\cite{1911Onnes,1957BardeenPhysRev}, the fractional quantum Hall effect~\cite{1999StormerRMP}, or the existence of giant magnetoresistance~\cite{1988BaibichPRL,1989BinaschPRB}. On the phenomenological level, collectivity relies on the amplification of quantum effects caused by quantum statistics and mutual interactions of individual particles when their number becomes macroscopic. In this case, the complex microscopic description can be effectively replaced by a mesoscopic one that, on the one hand, ignores the exact knowledge of the individual particles while still partially accounting for their properties and interactions. This approach stands behind all approximate methods based on the mean-field theory~\cite{1996GeorgesRMP,1999DalfovoRMP,2003BenderRMP}, the density functional approach~\cite{2015JonesRMP}, or the renormalization group schemes~\cite{1998FisherRMP}.

The precise study of mesoscopic quantum systems can be very fruitful in itself~\cite{BookMurayama2001}. This has become particularly appealing in the last two decades due to the development of extremely precise experimental possibilities for preparing and controlling systems containing a small number of strongly interacting ultracold atoms~\cite{2008CheinetPRL,2011SerwaneScience,2013WenzScience}. Thanks to a very high tunability, in such systems it is possible not only to control the number of particles precisely but also the strength of the interactions between them, the geometry of the external potential,  and the effective dimensionality. As a result, a remarkable avenue to explore non-classical correlations in mesoscopic-sized systems has been opened~\cite{2012BlumeRPP,2016ZinnerEPJWoC,2019SowinskiRPP}. As examples among others, let us mention here such phenomena as the spontaneous emerging of pairing in mesoscopic systems~\cite{2022HoltenNature,2024LairdSciPost}, or the emergence of Pauli crystals~\cite{2016GajdaEPL,2021HoltenPRL}.

The experimental study of ultracold gases has opened up many new possibilities for discovering previously not well-known quantum phenomena. This is due to the possibilities offered by the rich internal structure of the atoms. In particular, the existence of metastable atomic states with high total spin allows one to study exotic mixtures of quantum particles obeying various quantum statistics and experiencing highly non-trivial couplings between spin and spatial degrees of freedom~\cite{2014CazalillaRPP,2014PaganoNatPhys,2020SonderhouseNaturePhys}. Their properties in reduced dimensionality have become the subject of intensive research at many levels~\cite{2008LiaPRA,2009JiangEPL,2010GuanPRA,2011LeeJPA,2012KuhnNJP,2012GuanPRA,2016JiangJPA,2017LairdPRA,2020DobrzynieckiAdvQT,2024TaiPRR,2025SilvaPRA,2025TranNJP}, but their collective behavior induced by interactions in the few-body regime still requires systematic investigation. Moreover, due to the first spectacular experiments on multi-component few-body systems~\cite{2008OttensteinPRL,2023SchumacherARXiV}, a deeper theoretical understanding of systems hosting more than two components became very necessary. In our work, we take the first step in this direction.

In this work, we analyze a mesoscopic fermionic mixture consisting of three distinguishable components. Our aim is to test whether any collective effects may appear in the case of a larger number of components, which are not present in binary mixtures. In order to make these studies well-defined, reasonably comprehensive, and also going beyond extensively studied balanced mixtures, we consider the smallest possible extension of a two-component fermionic mixture by introducing additional interaction with a single third-component particle. We aim to investigate if and how interactions with this impurity change the internal properties of the mixture. Moreover, we inspect non-classical correlations that this may induce. In this way, we also link our work to the well-studied problem of a single impurity interacting with several spin-polarized fermions~\cite{2013WenzScience,2022WlodzynskiPRA}.

Our work is organized as follows. In Sec.~\ref{Section2} we introduce the theoretical model of the system studied and we describe its Hamiltonian in the convenient formalism. Then, in Sec.~\ref{Section3} we briefly explain the exact diagonalization method in the energy cut-off framework. In Sec.~\ref{Section4} we analyze single- and two-particle density distributions and we identify a specific transition in the many-body ground state of the system induced by interactions with the impurity. In Sec.~\ref{Section5} we study this transition in detail and we show that it has intriguing universal properties that do not depend essentially on either the number of particles in the system or the internal interactions in the mixture. Finally, we conclude in Sec.~\ref{Section6}.

\section{The system} \label{Section2}
In our work, we focus on the simplest possible extension of the two-component mixture of fermions to scenarios where interactions with an additional third-component particle (impurity) are present. We assume that all the particles have the same mass $m$, they are confined in the same one-dimensional parabolic trap of frequency $\Omega$, and they mutually interact via zero-range contact interactions in the $s$-wave channel. This implies, of course, that, due to the Pauli exclusion principle, only interactions between particles from different components do not vanish. Since in our considerations the third component is essentially distinguished, we write the Hamiltonian of the system conveniently as
\begin{align} \label{Hamiltonian}
\hat{\cal H} = \hat{\cal H}_{AB}+\hat{\cal H}_C + \hat{\cal H}_I.
\end{align}
In this sum, the Hamiltonian $\hat{\cal H}_{AB}$ describes the two-component mixture of fermions and in the second quantization reads
\begin{subequations}
\begin{align}
\hat{\cal H}_{AB} &= \int\!\mathrm{d}x\sum_{\sigma}\,\hat\Psi_\sigma^\dagger(x)\left[-\frac{\hbar^2}{2m}\frac{\mathrm{d}^2}{\mathrm{d}x^2}+\frac{m\Omega^2}{2}x^2\right]\hat\Psi_\sigma(x)  \nonumber \\
&+g_{0}\sqrt{\frac{\hbar^3\Omega}{m}}\int\!\mathrm{d}x\,\hat\Psi^\dagger_B(x)\hat\Psi^\dagger_A(x)\hat\Psi_A(x)\hat\Psi_B(x).
\end{align}
Here $\sigma\in\{A,B\}$ enumerates mixture's components while $\hat\Psi_\sigma(x)$ is a field operator annihilating $\sigma$-component particle at position $x$. These operators obey fermionic anti-comutation relations $\{\Psi_\sigma(x),\Psi^\dagger_{\sigma'}(x')\}= \delta_{\sigma\sigma'}\delta(x-x')$ and $\{\Psi_\sigma(x),\Psi_{\sigma'}(x')\}= 0$. The dimensionless parameter $g_0$ quantifies an effective interaction strength between components. 

Contrary, the Hamiltonian $\hat{\cal H}_C$ describes a third-component particle confined in the same trap. In the first quantization, it is represented by the differential operator of the form 
\begin{align} \label{HamiltonianC}
\hat{\cal H}_C = -\frac{\hbar^2}{2m}\frac{\mathrm{d}^2}{\mathrm{d}z^2}+\frac{m\Omega^2}{2}z^2,
\end{align}
where $z$ is the position of the particle.

Finally, the Hamiltonian $\hat{\cal H}_I$ is responsible for interactions between the two-component mixture and the third particle. It can be written as
\begin{align}
\hat{\cal H}_I = g \sqrt{\frac{\hbar^3\Omega}{m}}\int\!\mathrm{d}x\sum_\sigma\,\hat\Psi^\dagger_\sigma(x)\delta(x-z)\hat\Psi_\sigma(x),
\end{align}
\end{subequations}
where the dimensionless parameter $g$ is the effective interaction strength. Note that we assumed here that interactions with the third-component particle are the same for both components. Generalization to non-symmetric interactions is straightforward but beyond the scope of this work. We treat both interaction strengths $g_0$ and $g$ as independent parameters. It is justified from an experimental point since, in principle, they depend differently on the external magnetic field~\cite{2010ChinRMP}, the shape of the trapping potential in the transverse direction~\cite{1998OlshaniiPRL}, and the channel of interactions determined by the total spin of colliding particles~\cite{2017XuAnnPhys}. 

Let us also underline that in our approach we intentionally use different formalisms for different components, {\it i.e.}, the two-component mixture is described in the second quantization, which automatically takes into account the fermionic statistics, while the third-component particle is described within the first quantization emphasizing that there is only one particle there. In this notation, the total Hamiltonian $\hat{\cal H}$ acts in the Hilbert space being a tensor product of a single-particle Hilbert space of the third particle and a Fock space in the occupation number representation of the two-component mixture. This formalism has previously been used successfully to describe spinless fermions interacting with single impurity~\cite{2022WlodzynskiPRA}.

In the Hamiltonian \eqref{Hamiltonian} we already assumed many limiting simplifications: one-dimensionality, an equal mass of all particles, the same parabolic trap, zero-range $s$-wave interactions, or symmetric interaction with the third component. Still, however, there is a huge flexibility related to the number of particles in components $A$ and $B$. Fortunately, the Hamiltonian \eqref{Hamiltonian} commutes independently with operators of the total number of particles $\hat N_\sigma=\int\!\mathrm{d}x\, \hat\Psi^\dagger_\sigma(x)\hat\Psi_\sigma(x)$. It means that its properties can be studied independently in the subspaces of a well-defined number of particles. It is known that in the case of two-component mixtures (in our case, equivalent to setting $g=0$), measurable properties of the system strongly depend on the particle-number imbalance. This is related to a specific competition between the external potential constantly compressing the cloud of all particles, the quantum statistics acting inside fermionic components acting as quantum repelling pressure, and the inter-component interactions tending to spatially separate different components. To make our analysis as clear as possible, in this work, we focus only on symmetric mixtures having the same number of particles in both components, $N=N_A=N_B$. This assumption, together with the assumption of symmetric interactions with the third-component particle, leads directly to a full symmetry in the mixture sector. This symmetry is manifested formally by the fact that any eigenstate of the many-body Hamiltonian \eqref{Hamiltonian} is invariant under the interchange of A and B components. 

\section{The Method} \label{Section3}
To find the many-body ground state of the Hamiltonian \eqref{Hamiltonian} we adopt the numerically exact diagonalization technique. To make it feasible and accurate, we need to build an appropriately tuned many-body basis in which the Hamiltonian is represented and diagonalized. Since all particles are confined in the same external parabolic potential, it is convenient to choose the eigenstates of the harmonic oscillator as a single-particle basis for all components. They have a known textbook form
\begin{equation}
\varphi_n(z) = \left(2^n n!\sqrt{\pi a_0}\right)^{-1/2}\,\mathbf{H}_n\!\left(\frac{z}{a_0}\right)\mathrm{exp}\left(-\frac{z^2}{2a_0^2}\right),
\end{equation}
where $\mathbf{H}_n(\cdot)$ is $n$-th Hermite polynomial and $a_0=\sqrt{\hbar/m\Omega}$ is a natural unit of length in the problem. These functions are eigenfunctions of the operator \eqref{HamiltonianC} with eigenvalue $E_n=(n+1/2)\hbar\Omega$.  In this way one can decompose field operators  and introduce corresponding annihilation operators as
\begin{align}
\hat{\Psi}_A(x) =\sum_i \hat{a}_i\varphi_i(x), \qquad  \hat{\Psi}_B(x) =\sum_i \hat{b}_i\varphi_i(x).
\end{align}
Of course, they fulfill natural anticommutation fermionic relations. This gives us a natural path to introduce a basis in the many-body Hilbert space. As mentioned already, in our notation, the basis vectors have the form of a tensor product
\begin{equation} \label{FockState}
|\vec{\alpha},\vec{\beta},n\rangle = |\phi_n\rangle \otimes\left( a^\dagger_{\alpha_1}\cdots a^\dagger_{\alpha_N}b^\dagger_{\beta_1}\cdots b^\dagger_{\beta_N}|\mathtt{vac}\rangle\right),
\end{equation}
where $|\phi_n\rangle$ is a state of the third-component particle occupying $n$-th single-particle orbital, while two integer vectors $\vec{\alpha}=(\alpha_1,\ldots,\alpha_N)$ and $\vec{\beta}=(\beta_1,\ldots,\beta_N)$ encode single-particle orbitals occupied by individual fermions from components $A$ and $B$, respectively. Indistinguishability and fermionic nature of particles are guaranteed by imposing additional constraints of the form $\alpha_1>\alpha_2>\ldots>\alpha_N$ and $\beta_1>\beta_2>\ldots>\beta_N$. States \eqref{FockState} are obviously eigenstates of the non-interacting Hamiltonian ($g_0=g=0$) and have energies $E_{\vec\alpha,\vec\beta,n}=\hbar\Omega \left[(2N+1)/2 + n + \sum_i (\alpha_i + \beta_i)\right]$. The non-interacting ground state has energy $E_0=\hbar\Omega(N^2+1/2)$. 

\begin{figure}
\includegraphics[width=\linewidth]{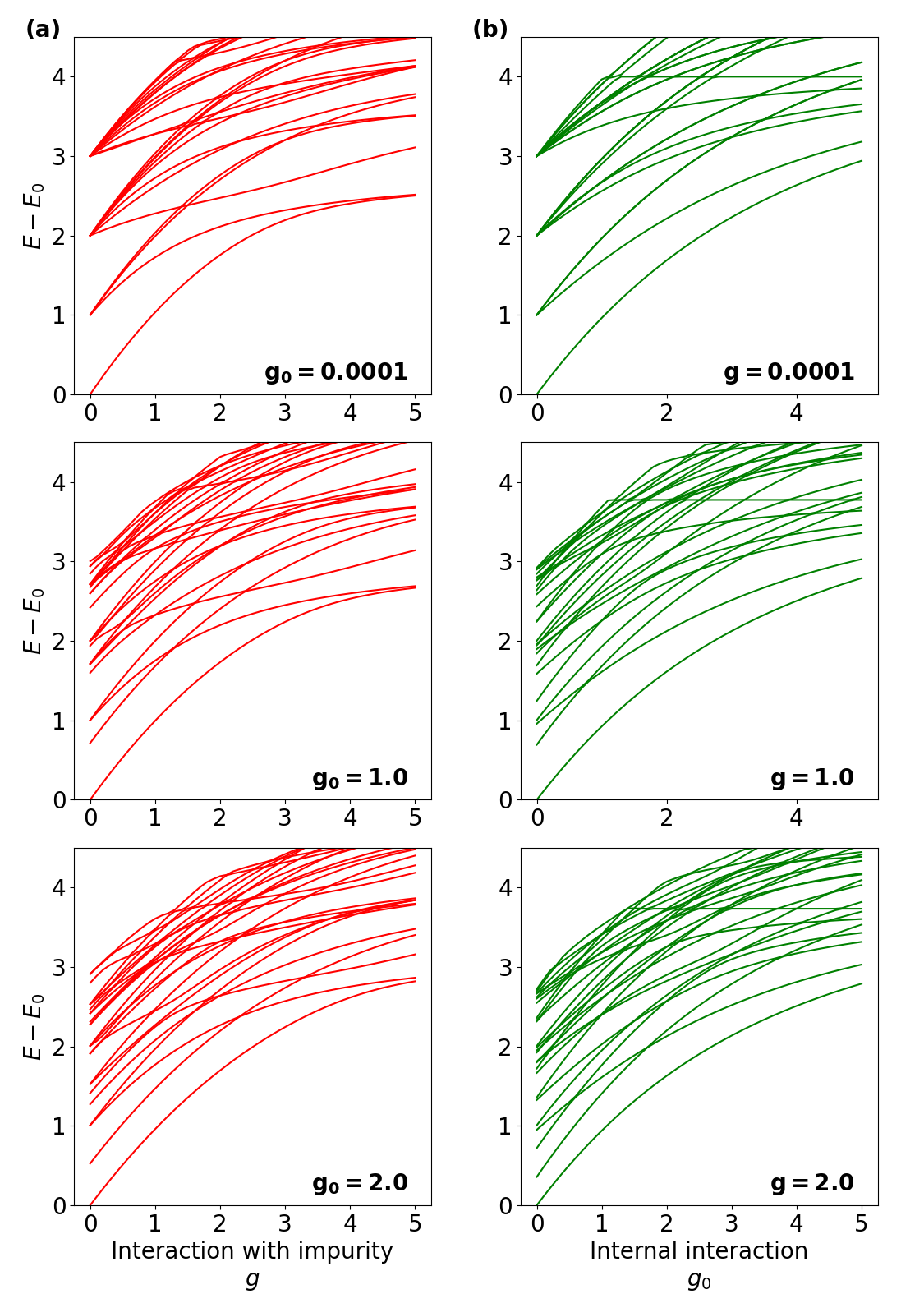}
\caption{Low-energy spectrum of the many-body Hamiltonian \eqref{Hamiltonian} obtained for a system containing an impurity interacting with the two-component mixture of four particles ($N=2$). {\bf (a)} Eigenenergies as functions of interactions with the impurity $g$ for a fixed internal interaction $g_0$. {\bf (b)} Eigenenergies as functions of internal interactions $g_0$ for a fixed interactions with the impurity $g$. In all plots, energies are expressed in natural units of energy $\hbar\Omega$, and for clarity, we shift them by the ground-state energy at $g=0$ for (a) and $g_0=0$ for (b), respectively.
\label{Fig1}}
\end{figure}

To diagonalize the Hamiltonian numerically, one needs to cut the basis. In our approach, we do not perform simple cuts on single-particle excitations, but rather we limit the many-body basis to states having total (non-interacting) energy $E_{\vec\alpha\vec\beta n}$ not larger than some fixed energy limit $E_C$. In this way, we do not exclude from the description low-energy states having only a few highly excited particles, while we neglect high-energy states with many low-excited particles. It was shown in previous studies that such a method gives a much better approximation of interacting many-body states with the same usage of numerical resources~\cite{1998HaugsetPRA,2018PlodzienARXIV,2019ChrostowskiAPPA,2022RojoPRA}. In our case, the determination of the appropriate value of cut-off energies $E_C$ for specific systems considered is based on a convergence test of the many-body ground state. A detailed description of this procedure and the cut-off energy values used is provided in the Appendix~\ref{appendix}. Following this procedure, for a given number of particles $N$ and interaction strengths $g_0$ and $g$, we represent the Hamiltonian \eqref{Hamiltonian} in the truncated many-body basis and diagonalize with the implicitly-restarted Arnoldi-Lanczos algorithm~\cite{1998BookARPACK}. As a result, we obtain the lowest many-body eigenstates represented as a superposition of basis states $|{\vec\alpha,\vec\beta,n}\rangle$ and their energies.  In this way, we have particular access to a whole structure of the many-body ground state $|\mathtt{G}\rangle$. We present exemplary spectra for the system of five particles ($N=2$) as a function of the interaction strengths in Fig.~\ref{Fig1}. Firstly, one notes that the spectrum appears to display features that could be indicative of quantum chaos, although confirming this would require more detailed analysis. This type of analysis goes significantly beyond the scope of this work, and we leave it for future studies. Importantly for our studies, we note that the ground state is isolated for any interaction parameters $g_0$ and $g$, and degeneracy may occur only in the limit of infinite repulsions. This is a general observation valid for any number of particles we consider, and it means that any observed structural changes in the ground state do not occur due to the intersection of states in the spectrum and are an inherent property of the ground state itself. 

Let us mention here that in the case of harmonic confinement, further optimization techniques of the many-body basis are possible~\cite{2015VolosnievEPJST,2017BellottiEPJD,2018KoscikPLA,2020KoscikFBS}. They are vital when large interaction strengths are considered. In the cases of relatively weak interactions studied in our work, this kind of optimization has limited importance. Therefore, we do not introduce them to our scheme.

\section{Impact of the impurity} \label{Section4}
\begin{figure}
\includegraphics[width=\linewidth]{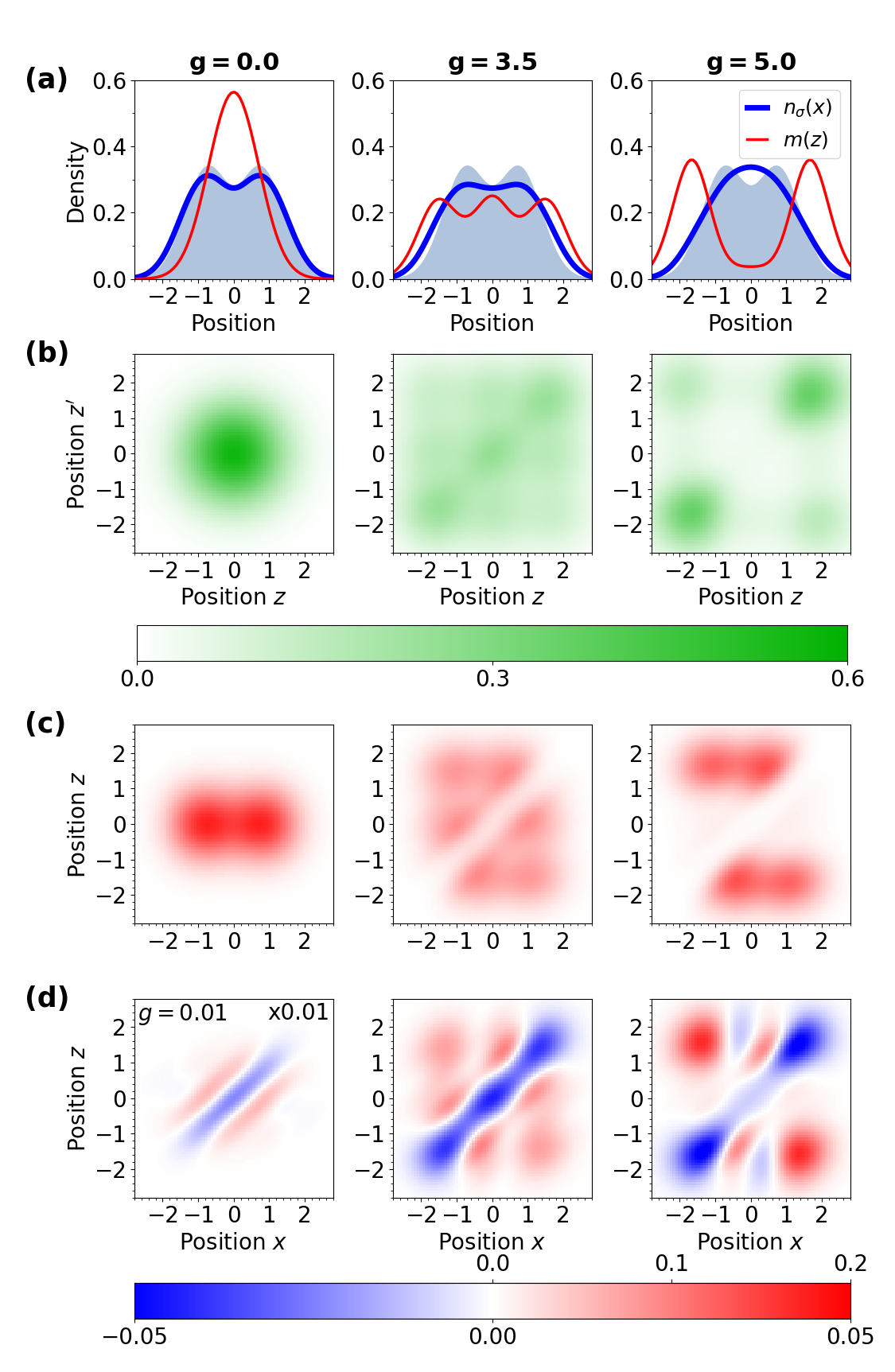}
\caption{Impact of interactions with the third-component particle on the two-component mixture containing four particles ($N=2$). Different columns correspond to different values of interaction strength with the impurity $g$. In all plots, inter-component interaction strength is fixed, $g_0=1$. {\bf (a)} Single-particle density profiles of the mixture's components $n_\sigma(x)$ (thick blue) and the third-component particle $m(x)$ (thin red). For a better comparison, the non-interacting distribution of the mixture's components is additionally displayed with a light blue shadow. {\bf (b)} Single-particle reduced density matrix of the impurity $\rho(x,x')$. {\bf (c)} Two-particle density profile $\mu_{\sigma}(x,z)$. {\bf (d)} Distribution of correlations ${\cal G}_\sigma(x,z)$ between impurity and particle from the component $\sigma$. Note that in the left plot, we consider the interaction strength $g$ slightly positive ($g=0.01$) since for vanishing interactions the distribution vanishes. In this case, the distribution intensity is also scaled by a factor $10^2$ to make it visible. The upper and lower values on the color bar refer to plots  (c) and (d), respectively. In all plots, positions are given in units of $a_0$. In panels (a) and (b), densities are expressed in units of $a_0^{-1}$, while in panels (c)-(d), in units of $a_0^{-2}$.
\label{Fig2}}
\end{figure}  
\begin{figure}
\includegraphics[width=\linewidth]{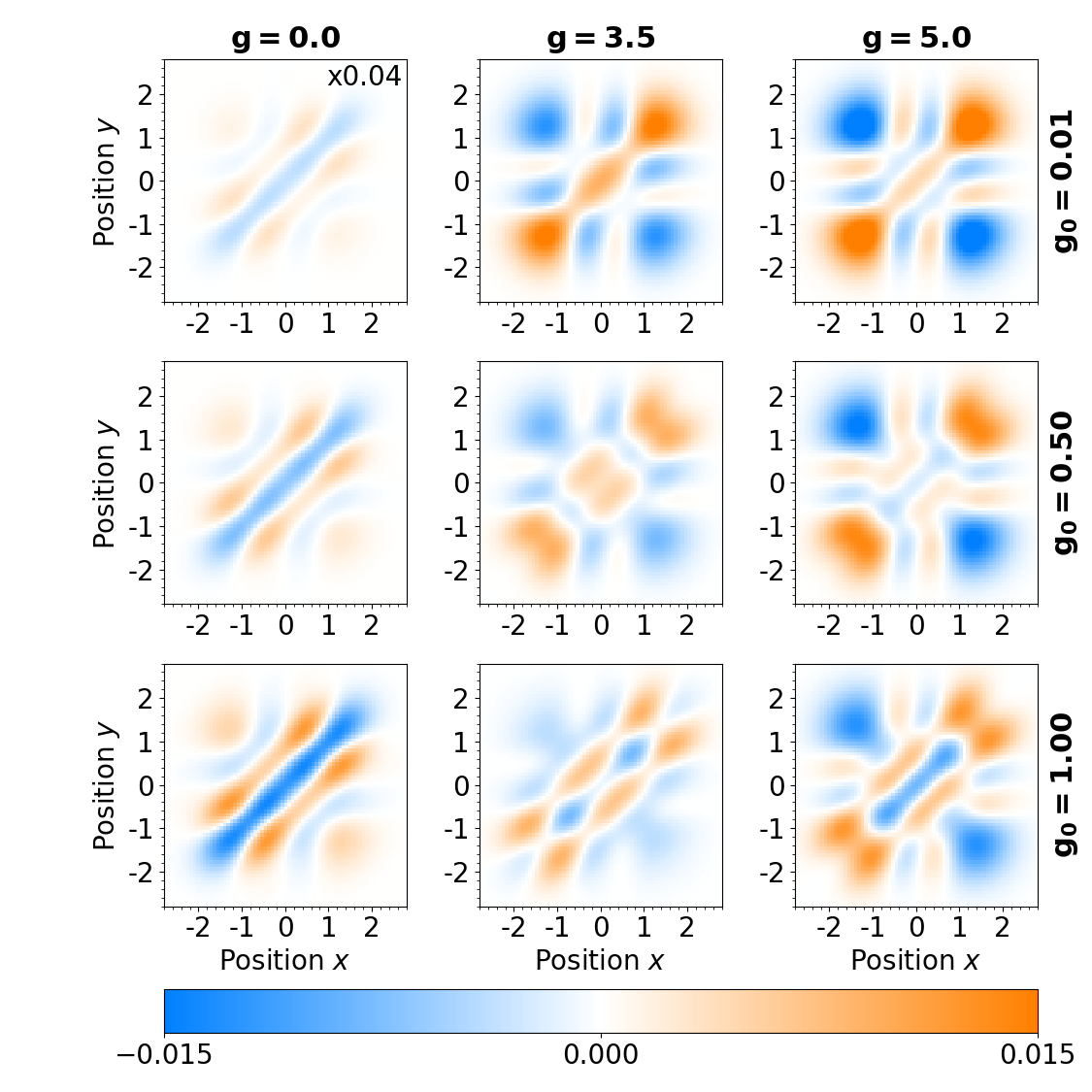}
\caption{Distribution of internal two-particle correlations ${\cal K}(x,y)$ in the system containing impurity interacting with the two-component mixture containing four particles ($N=2$) for different internal interaction strengths $g_0$ and interactions with the impurity $g$. For any fixed interaction $g_0$, along with increasing repulsion $g$, the transition between two different regimes is visible. In all plots, positions are expressed in units of $a_0$ while densities are in units of $a_0^{-2}$. \label{Fig3}}
\end{figure}

In principle, the impact of interactions with an additional impurity on the system can be analyzed from two complementary perspectives, depending on whether one is interested in the external or internal behavior of the mixture. First, let us discuss the external features when interactions with the impurity are present. The simplest quantities that feel an influence of additional interaction are single-particle density distributions $n_\sigma(x)$ of mixture components and $m(x)$ of the impurity. They are straightforwardly defined as
\begin{subequations}
\begin{align}
n_\sigma(x) &= \langle \mathtt{G}|\mathbb{1}\otimes\hat\Psi_\sigma^\dagger(x)\hat\Psi_\sigma(x)|\mathtt{G}\rangle, \\
m(x) &=  \langle\mathtt{G}| \hat{\mathbb{P}}_x \otimes \mathbb{1}|\mathtt{G}\rangle,
\end{align}
\end{subequations}
where $\hat{\mathbb{P}}_x=|x\rangle\langle x|$ projects to an eigenstate $|x\rangle$ of position operator. Of course, due to the exchange symmetry $A\leftrightarrow B$ explained before, both density distributions $n_A(x)$ and $n_B(x)$ are exactly the same. In Fig.~\ref{Fig2}(a), we present these distributions (thick blue and thin red lines) for the system of five particles ($N=2$) for different interaction strengths $g$, keeping fixed internal interaction $g_0=1$. For better comparison, we additionally display with the shadowed area the non-interacting ($g_0=g=0$) distribution $n_\sigma(x)$. It is clear that along with increasing $g$, the distribution of the impurity $m(x)$ significantly changes its character and for sufficiently large repulsions it is split and pushed out from the center where the mixture is concentrated. In contrast, distributions of mixture components $n_\sigma(x)$ are only slightly affected. Since the impact of interactions on the impurity is significant, it is also worth analyzing the off-diagonal order of the separated particle. This is encoded directly in the single-particle reduced density matrix, which in our notation can be written as 
\begin{equation}
\rho(x,x') = \langle\mathtt{G}| \left(|x\rangle\langle x'| \otimes \mathbb{1}\right)|\mathtt{G}\rangle.
\end{equation}
Obviously, a diagonal part of this quantity is equal to a single-particle density, $m(x)=\rho(x,x)$. As clearly seen in Fig.~\ref{Fig2}(b), the reduced density matrix undergoes a substantial change when the system is driven between the two regimes. In particular, distinct isolated enhancements appear at the end of the diagonal and anti-diagonal. They are a direct consequence of splitting and pulling apart the probability distribution of a single particle. 

This behavior of single-particle properties can be investigated further by considering the two-particle density distribution 
\begin{equation}
\mu_\sigma(x,z) =\langle \mathtt{G}|\hat{\mathbb{P}}_z\otimes\hat\Psi_\sigma^\dagger(x)\hat\Psi_\sigma(x)|\mathtt{G}\rangle
\end{equation}
which encodes the joint probability of finding the impurity at position $z$ and one of the particles from the component $\sigma$ at position $x$. The distribution for the same parameters of the system is shown in Fig.~\ref{Fig2}(c). In addition to the spatial separation already deduced from single-particle distributions, the two-particle density clearly indicates that some correlations between mixture and impurity are developed. Most significantly, they are visible along the diagonal direction $z=x$, {\it i.e.}, the probability of finding both particles at the same position quickly vanishes with increasing interaction $g$. Of course, this effect cannot be captured by single-particle distributions and is not present in the direct product $n_A(x)m(z)$. To make it evident, we additionally display in Fig.~\ref{Fig2}(d) the distribution of correlations defined as
\begin{equation} \label{CorrNoiseG}
{\cal G}_\sigma(x,z) = \mu_\sigma(x,z)-n_\sigma(x)m(z).
\end{equation}
Its negative (positive) value indicates that the actual probability is smaller (higher) than the probability that would result from the assumption of complete independence of the components.

From the perspective of the internal properties of the mixture, the description is slightly simpler. As noticed already, interactions with the impurity do not significantly influence the single-particle distribution of the mixture. It turns out that this is also the case when one considers the inter-component two-particle distribution
\begin{equation}
\nu(x,y) =\langle \mathtt{G}|\hat{\mathbb{1}}\otimes\hat\Psi_A^\dagger(x)\hat\Psi_A(x)\hat\Psi_B^\dagger(y)\hat\Psi_B(y)|\mathtt{G}\rangle.
\end{equation}
The impact of the impurity is slightly noticeable when the distribution of two-particle correlations is analyzed. It is defined analogously to correlations \eqref{CorrNoiseG} as
\begin{equation}
\label{CorrNoiseK}
{\cal K}(x,y) = \nu(x,y)-n_A(x)n_B(y).
\end{equation}
In Fig.~\ref{Fig3} we show this distribution for different values of interactions $g_0$ and $g$ for the same system of five particles ($N=2$). The magnitude of correlations is very small when compared to the magnitude of correlations in ${\cal G}_\sigma(x,z)$ and the transition is not as clearly visible as in Fig.~\ref{Fig2}. However, one can still notice that correlations along the anti-diagonal (especially far from the center) change their character when the system undergoes the transition. Nonetheless, it is true that from the perspective of internal correlations within the mixture, the structural transition is not very clearly manifested.

Up to now, we have presented illustrative results for a system containing five particles (two particles in each mixture's component and the impurity). For completeness, we performed analogous calculations for systems with a larger number of particles in the mixture (up to $N=5$). They are exposed in the Supplemental Material~\cite{zenodo}. On a quantitative level, the results obtained are very similar in the sense that the observed structural change in the many-body ground state is universal. Namely, for a given number of particles and fixed internal interaction $g_0$, one finds such an interaction strength $g$ at which distributions change rapidly. Particularly,  the single-particle density $m(x)$ is split and pushed to the edges while the distribution of internal correlations ${\cal K}(x,y)$ changes its nature.

\begin{figure}
\centering
\includegraphics[width=\linewidth]{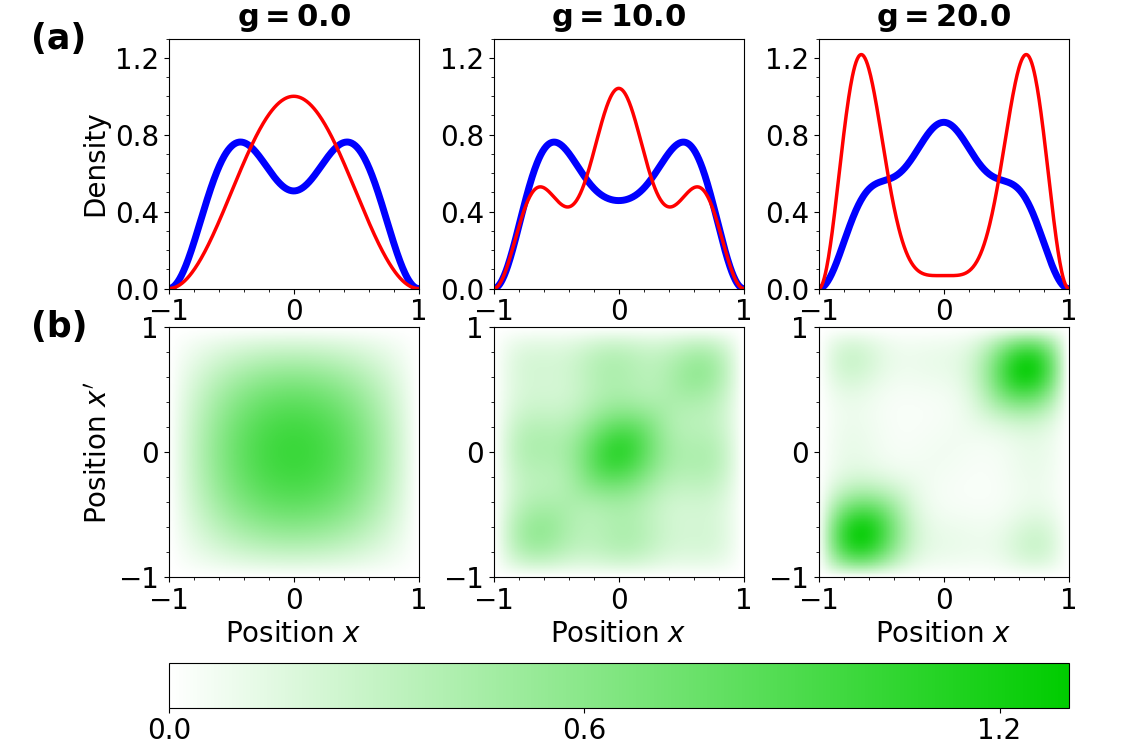}
\caption{Impact of interactions with the third-component particle on the two-component mixture containing four particles ($N=2$) when the system is confined in a rectangular potential well \eqref{HamiltonianWell}. Different columns correspond to different values of interaction strength with the impurity $g$. In all plots, inter-component interaction strength is fixed, $g_0=1$. {\bf (a)} Single-particle density profiles of the mixture's components $n_\sigma(x)$ (thick blue) and the third-component particle $m(x)$ (thin red). {\bf (b)} Single-particle reduced density matrix of the impurity $\rho(x,x')$. Note that for sufficiently large interaction $g$, the impurity is separated similarly as in the case of a harmonic confinement. In all plots, positions are expressed in the width of the well $L$, while densities are in $L^{-1}$. } \label{Fig5}
\end{figure}
From phenomenological argumentation, some kind of separation of the components is suspected. When interactions are strong enough, the single-particle excitations become energetically favourable over the interaction energy cost. Then, to reduce the overlap of the densities, the system excites particles. It should be noted, however, that the separation observed here is substantially different from separations observed in systems of single-component fermions interacting with the impurity~\cite{2014LindgrenNJP}. It was shown that in those systems, the impurity remains in the middle of the trap while the fermionic component is (only slightly) split and pushed out from the center. In our case, although repulsive interactions in the fermionic mixture are present, the distribution of the impurity is split and driven to the edges. As noted in the literature, in the case of the single-component Fermi sea, the existence of this kind of separation requires a noticeable mass difference between impurity and other particles~\cite{2016PecakNJP,2022WlodzynskiPRA}. Still, however, there is a substantial difference between the scenario studied here and two-component mass-imbalanced systems. To demonstrate this, it is sufficient to examine how the separation depends on the shape of the external potential, when the Hamiltonian \eqref{Hamiltonian} has the following form
\begin{subequations} \label{HamiltonianWell}
\begin{align}
\hat{\cal H}_{AB} &= \int_{-L}^L\!\mathrm{d}x\sum_{\sigma}\,\hat\Psi_\sigma^\dagger(x)\left(-\frac{\hbar^2}{2m}\frac{\mathrm{d}^2}{\mathrm{d}x^2}\right)\hat\Psi_\sigma(x)  \nonumber \\
&+g_{0}\frac{\hbar^2}{mL}\int_{-L}^L\!\mathrm{d}x\,\hat\Psi^\dagger_B(x)\hat\Psi^\dagger_A(x)\hat\Psi_A(x)\hat\Psi_B(x),\\
\hat{\cal H}_C &= -\frac{\hbar^2}{2m}\frac{\mathrm{d}^2}{\mathrm{d}z^2} + V(z), \\
\hat{\cal H}_I &= g \frac{\hbar^2}{mL}\int_{-L}^L\!\mathrm{d}x\sum_\sigma\,\hat\Psi^\dagger_\sigma(x)\delta(x-z)\hat\Psi_\sigma(x),
\end{align}
\end{subequations} 
where $V(z) = 0$ for $|z|<L$ and $V(z)\rightarrow \infty$ otherwise. It is known that in the case of two-component systems with different masses, the lighter component undergoes separation in the harmonic potential, but in a rectangular potential well, the heavier component is separated~\cite{2016PecakPRA}. In Fig.~\ref{Fig5}, we show that in the case of a two-component mixture interacting with the impurity described with Hamiltonian \eqref{HamiltonianWell}, there is no such change induced by the shape of the potential -- the impurity density is again subjected to separation. 

\begin{figure}
\includegraphics[width=\linewidth]{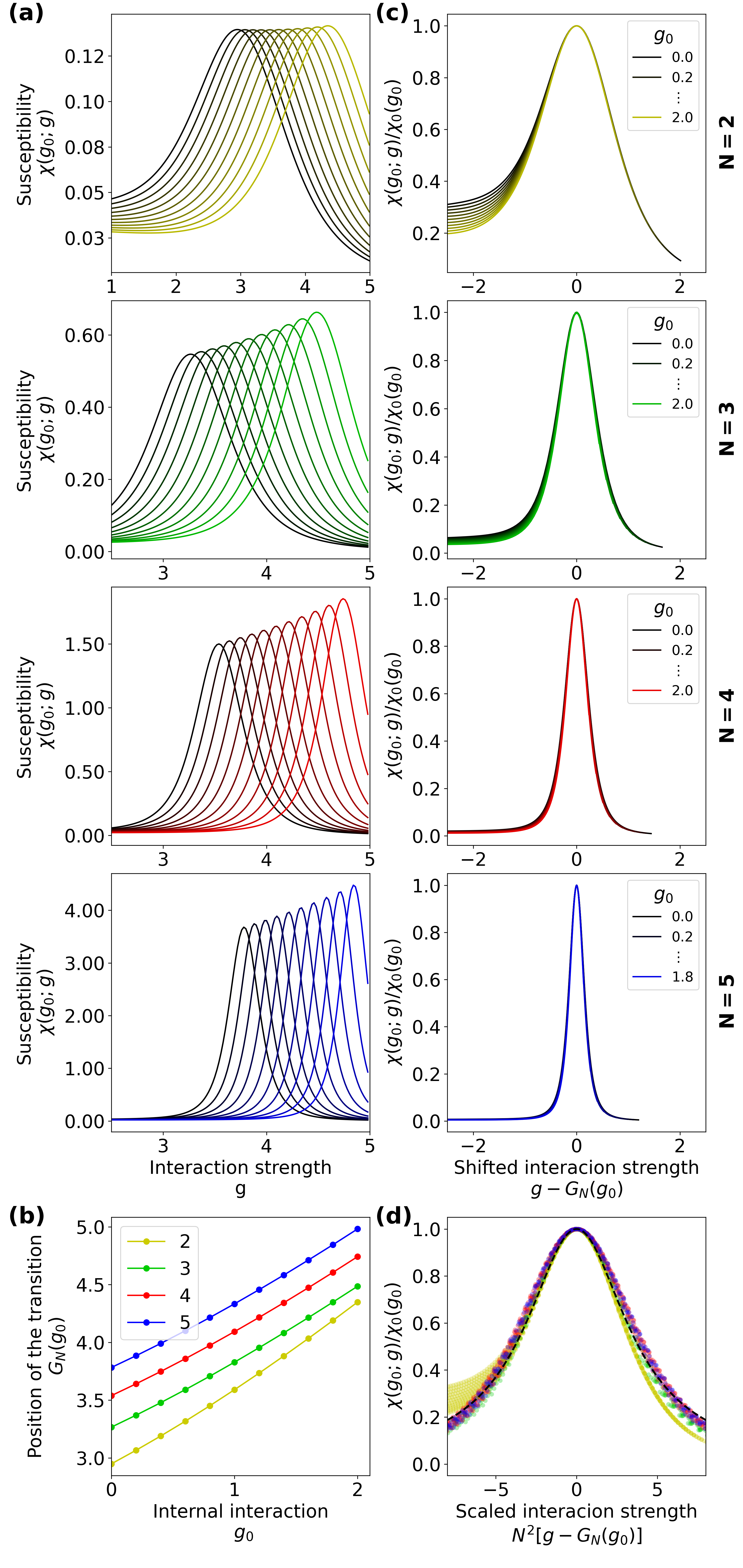}
\caption{Properties of the fidelity susceptibility for different numbers of particles in the system. {\bf (a)}  The fidelity susceptibility $\chi(g_0;g)$ as a function of interaction $g$ for different internal interactions $g_0$ (legend in panel (c)). {\bf (b)} Position of the transition $G_N(g_0)$ as a function of internal interaction $g_0$. {\bf (c)} Susceptibilities from the left plot after rescaling in magnitude by $\chi_0(g_0)$ and centering around the transition interaction $G_N(g_0)$. All plots collapse to the same universal curve, signalling the specific universality of the transition. {\bf (d)} Numerical collapse to the universal curve $\Phi$ (dashed black) of all the fidelity susceptibility curves (different points and colors) after applying appropriate scaling~\eqref{FidelFinal}. Note that deviations from the universality are diminished when the number of particles in the mixture increases.\label{Fig4}}
\end{figure}

\section{The structural transition} \label{Section5}
As explained already, the system studied experiences some substantial transition in its many-body ground state, and a precise quantitative study of its nature is needed. 
To get a better understanding of this transition, we utilize additional quantitative tools. Since the method of exact diagonalization gives direct access to the structure of the many-body ground state, one can calculate the fidelity between ground states obtained for different interaction strengths. When internal interaction $g_0$ is fixed, the fidelity is defined as

\begin{equation}
{\cal F}(g_0;g,g') = |\langle \mathtt{G}(g_0,g)|\mathtt{G}(g_0,g')\rangle|^2,
\end{equation}

where we explicitly mark that the many-body ground state $|\mathtt{G}(g_0,g)\rangle$ is calculated for interactions $g_0$ and $g$, respectively. Since for $g=g'$ the fidelity is extreme and equal to 1, one expands it naturally into the series
\begin{equation}
{\cal F}(g_0;g,g') = 1 - \chi(g_0;g)(g'-g)^2 + \ldots
\end{equation}
In this expansion, $\chi(g_0;g)=\partial^2_{g'} {\cal F}(g_0;g,g')|_{g'=g}$ is the fidelity susceptibility which quantifies the resistivity of the ground state on infinitesimal change of interaction strength~\cite{2007YouPRE,2015WangPRX}. Its large amplitude for a given value of $g$ signals that at this point the ground state of the system rapidly changes with $g$, {\it i.e.}, the state of the system quickly moves toward different regions of the Hilbert space. In our numerical approach, we find susceptibility $\chi(g_0;g)$ straightforwardly by using a three-point approximation of a second derivative with a small step $\delta g=g'-g=10^{-4}$.

In Fig.~\ref{Fig4}(a), we present the susceptibility $\chi(g_0;g)$ as a function of interaction with the impurity $g$ for different internal interactions $g_0$. Successive rows correspond to varying numbers of particles in the mixture $N$. It is clear that the position of the transition $G_N(g_0)$, {\it i.e.}, interaction strength $g$ for which the susceptibility is maximal (Fig.~\ref{Fig4}(b)), depends monotonically on the internal interaction in the mixture $g_0$. The intensity of the transition (magnitude of the susceptibility) is evidently higher for systems with a larger number of particles. Contrary, the shape of the function $G_N(g)$ is not strongly affected by the number of particles.

One of the most interesting properties of the observed transition is its remarkable universality. It is easily recognized with an appropriate rescaling of the susceptibility function. To make it evident, in Fig.~\ref{Fig4}(c), we plot susceptibilities as a function of distance from the transition point $\tau=g-G_N(g_0)$ and normalized to its maximal value $\chi_0(g_0)=\chi(g_0;G_N(g_0))$. We find then that, in the vicinity of the transition, all the susceptibility curves obtained for a given number of particles $N$ follow a single universal distribution $\phi_N(\xi)$. Of course, with increasing distance from the transition point, the curves determined for different $g_0$ become increasingly different from the universal shape. Note, however, that these deviations are diminished when the number of particles in the mixture is increased. This observation suggests that for any internal interaction strength $g_0$ the susceptibility can be written as
\begin{equation}
\chi(g_0;g)= \chi_0(g_0)\cdot\phi_N\big(g-G_N(g_0)\big).
\end{equation}
In principle, the $N$-dependent universal curve $\phi_N(\xi)$ is different for different numbers of particles in the mixture. For example, it is visible that for a larger number of particles, it becomes significantly narrower, indicating that the transition is more rapid. We found, however, additional simple scaling between systems of different numbers of particles, which makes all the curves reasonably overlapping. Namely, all $N$-dependent functions $\phi_N(\xi)$ can be fused to a single universal curve $\Phi(\xi)$ by applying the scaling $\phi_N(\xi)=\Phi(N^2 \xi)$. In Fig.~\ref{Fig4}(d), we display all the fidelity susceptibilities from panel (a) after the aforementioned scaling and shifting. We see that all the transition plots follow almost the same curve, and the agreement is better for a larger number of particles. Consequently, we argue that any fidelity susceptibility curve can be derived straightforwardly from the universal function $\Phi(\xi)$ via the relation
\begin{equation} \label{FidelFinal}
\chi(g_0;g)= \chi_0(g_0)\cdot\Phi\Big(N^2\big(g-G_N(g_0)\big)\Big).
\end{equation}
The observed scaling with the number of particles in the mixture is the same as the scaling of the internal interaction energy in the two-component mixture. From this perspective, all susceptibility curves are identical. This also indirectly explains why the scaling found becomes better with an increase in the number of particles. 

Let us underline here that the shape of the universal curve $\Phi(\xi)$ is not known and probably cannot be deduced with any simple heuristic arguments. However, it turns out that it can be well approximated by a simple one-parameter Lorentzian shape of the form
\begin{equation}
\Phi(\xi) = \frac{\Gamma_0^2}{\xi^2 + \Gamma_0^2}.
\end{equation}
The numerical fit to all the data (for $N>2$) gives $\Gamma_0 \approx 3.84$ and the resulting shape is presented in Fig.~\ref{Fig4}(d) with a dashed black line. The observed universality of the structural transition and its simple scaling with particle number suggest that the transition is driven rather by a collective mechanism that is only weakly influenced by the microscopic properties of individual particles. 

Lastly, in light of our results, we intend to note that the universality just described is probably a feature for systems where the number of particles in the mixture is large. Indeed, as already mentioned, for smaller $N$, deviations from the universal curve become more pronounced. Moreover, we checked that for the smallest system (three particles, single particle in each component, $N=1$), discrepancies are so huge that claiming any universality is not justified.

\section{Conclusions} \label{Section6}
In our work, we extensively investigated the ground-state properties of the simplest three-component fermionic mixtures, {\it i.e.}, a balanced two-component mixture interacting additionally with a single impurity from the third flavor. By applying the method of exact diagonalization, we access not only the ground-state energy and density profiles of particular components but also the mutual correlations existing in the system. We observed that with increasing interaction with the impurity, a specific ground-state transformation occurs in the system, during which the third component (the impurity) is split and pushed to the edges of the system. Although the interaction strength at which the transition occurs depends on the number of particles and internal interactions in the mixture, it has many universal features. In particular, the shape of the susceptibility curve does not depend (after taking into account the scaling of interactions) on the internal parameters of the mixture.

Our work is the next step towards a better understanding of the properties of imbalanced multi-component mixtures of several strongly correlated ultracold fermions. The observed transition and its universality show that a detailed analysis of such systems can expose many of their interesting properties that have no counterparts in systems with fewer components. There are at least three natural directions for extending our work. One is to consider interactions of the impurity with balanced mixtures of more than two components. The other is to consider interactions with more than one impurity with a multi-component mixture and explore substantial differences with earlier results for a spinless Fermi sea~\cite{2019MistakidisNJP}. Thirdly, these directions should be simultaneously extended with an analysis of attractive interactions between all or only certain components of a multi-component system.

\section*{Data availability statement}
All numerical data presented in this paper and additional figures for different numbers of particles in the mixture are available online~\cite{zenodo}.

\section*{Acknowledgments}
This research was supported by the National Science Centre (NCN, Poland) within the OPUS project No.~2023/49/B/ST2/03744. For the purpose of Open Access, the authors have applied a CC-BY public copyright licence to any Author Accepted Manuscript version arising from this submission.

\appendix
\section{Convergence test}\label{appendix}
\begin{figure}
\centering
\includegraphics[width=\linewidth]{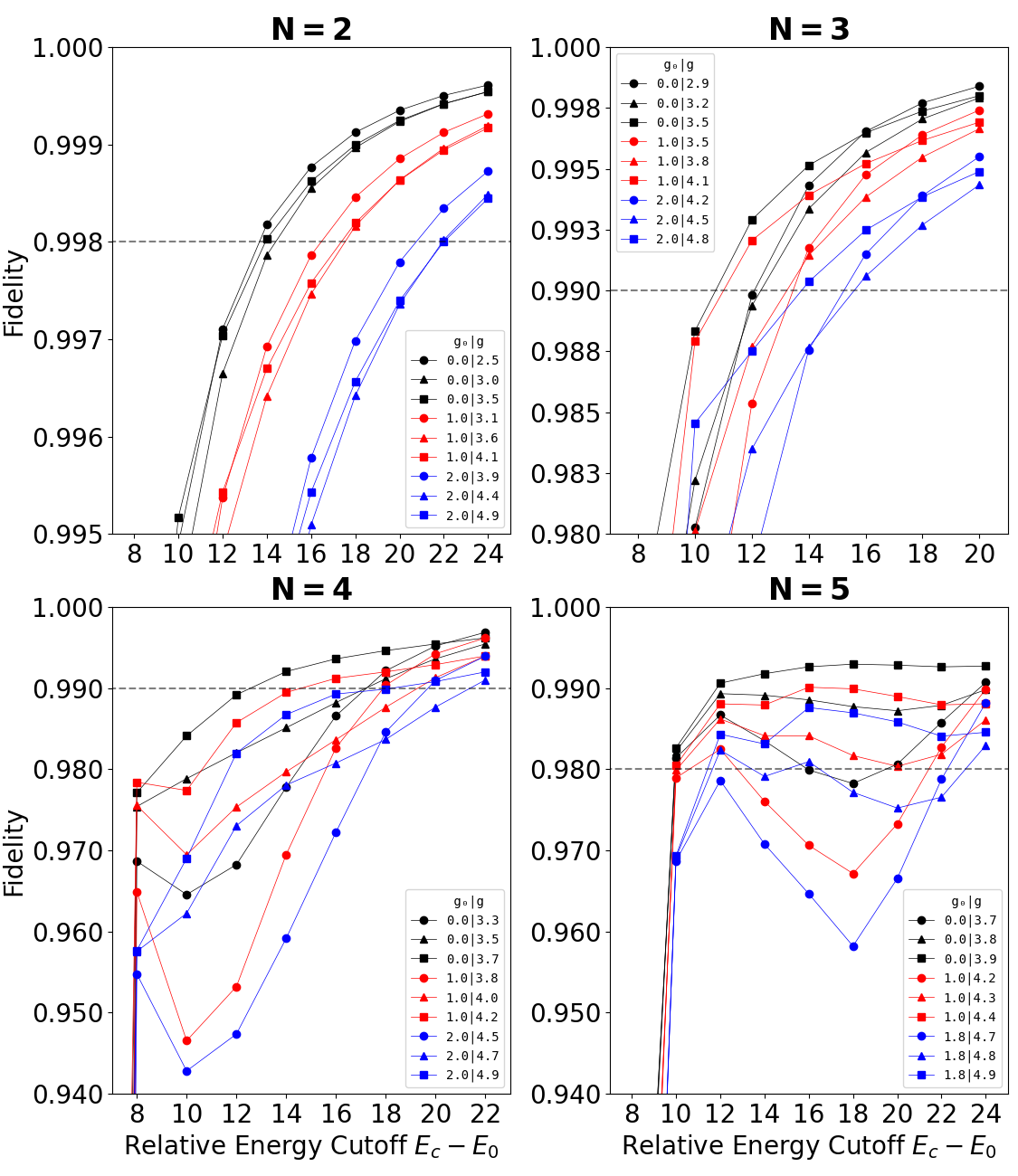}
\caption{Fidelity \eqref{FidCut} between ground states obtained for successively increased cut-offs $E_C$ (relatively to the non-interacting energy $E_0$), for different numbers of particles and different interaction strengths -- before ($\bigcirc$), after ($\triangle$), and close to ($\square$) the structural transition point. We find that for sufficiently large cut-off the fidelity progressively approaches unity and in all the cases studied is fairly larger than $98\%$. All energies are expressed in natural units of harmonic oscillator $\hbar\Omega$. \label{Fig6}}
\end{figure}

In our work, to determine the many-body ground state of the Hamiltonian \eqref{Hamiltonian}, we use the exact diagonalization approach. Its accuracy significantly depends on the energy cut-off $E_C$ used during the cropping of the many-body basis. For a given system, we determine a sufficient cut-off by calculating the fidelity between ground states $|\mathtt{G}(E_C)\rangle$ obtained for successively increased cut-offs (with step $2\hbar\Omega$ due to a mirror symmetry of the system)
\begin{equation} \label{FidCut}
\mathbf{F}(E_C) = |\langle \mathtt{G}(E_C)|\mathtt{G}(E_C-2\hbar\Omega)\rangle|^2.
\end{equation}
and observing its convergence to unity. In Fig.~\ref{Fig6} we display examples of this dependence for different numbers of particles and different interaction strengths. For clarity, instead of pure cut-off $E_C$, in all plots we use cut-off energy relative to the non-interacting energy $E_0=\hbar\Omega(N^2+1/2)$. Different colors and shapes correspond to different strengths of internal interactions $g_0$ and interactions with the impurity $g$, respectively. Particularly, squares ($\square$) represent the cases in the vicinity of the structural transition point, while circles ($\bigcirc$) and triangles ($\triangle$) represent situations on the opposite sides of this transition. In all the cases, we fairly see that the fidelity approaches unity with increasing cut-off. With a cut-off chosen such that the fidelity exceeds $98\%$, our numerical resources are sufficient to investigate the structural transition in systems containing up to 11 particles ($N=5$). Of course, for smaller systems, the threshold can be significantly improved within the same numerical resources. This allowed us to verify (for smaller systems) that further increasing the cut-off does not noticeably change our results. Therefore, we conclude that for the largest system studied ($N=5$), a threshold of $98\%$ is sufficient. Final cut-offs taken in our numerical approach and dimension ${\cal D}_\mathrm{Fock}$ of corresponding Fock spaces are given in Table~I.
\begin{table}\label{table1}
\begin{tabular}{@{\hskip 0.25cm}c@{\hskip 0.5cm}c@{\hskip 0.5cm}c@{\hskip 0.5cm}c@{\hskip 0.5cm}c@{\hskip 0.25cm}}
\hline \hline
$N$ & $E_C - E_0$ & $E_C$ & Threshold & ${\cal D}_\mathrm{Fock}$ \\
\hline
2 & 24 & 29 & 99.8\% & 35 399\\
3 & 20 & 30 & 99.0\% & 53 577\\
4 & 22 & 39 & 99.0\% & 186 113\\
5 & 24 & 50 & 98.0\% & 520 519\\
\hline \hline
\end{tabular}
\caption{Determined energy cut-off $E_C$ for different number of particles $N$. All the cut-offs meet the criterion of less than $2\%$ fidelity deviation between successive ground states. }
\end{table}

\bibliography{biblio}

\begin{thebibliography}{60}%
\makeatletter
\providecommand \@ifxundefined [1]{%
 \@ifx{#1\undefined}
}%
\providecommand \@ifnum [1]{%
 \ifnum #1\expandafter \@firstoftwo
 \else \expandafter \@secondoftwo
 \fi
}%
\providecommand \@ifx [1]{%
 \ifx #1\expandafter \@firstoftwo
 \else \expandafter \@secondoftwo
 \fi
}%
\providecommand \natexlab [1]{#1}%
\providecommand \enquote  [1]{``#1''}%
\providecommand \bibnamefont  [1]{#1}%
\providecommand \bibfnamefont [1]{#1}%
\providecommand \citenamefont [1]{#1}%
\providecommand \href@noop [0]{\@secondoftwo}%
\providecommand \href [0]{\begingroup \@sanitize@url \@href}%
\providecommand \@href[1]{\@@startlink{#1}\@@href}%
\providecommand \@@href[1]{\endgroup#1\@@endlink}%
\providecommand \@sanitize@url [0]{\catcode `\\12\catcode `\$12\catcode
  `\&12\catcode `\#12\catcode `\^12\catcode `\_12\catcode `\%12\relax}%
\providecommand \@@startlink[1]{}%
\providecommand \@@endlink[0]{}%
\providecommand \url  [0]{\begingroup\@sanitize@url \@url }%
\providecommand \@url [1]{\endgroup\@href {#1}{\urlprefix }}%
\providecommand \urlprefix  [0]{URL }%
\providecommand \Eprint [0]{\href }%
\providecommand \doibase [0]{http://dx.doi.org/}%
\providecommand \selectlanguage [0]{\@gobble}%
\providecommand \bibinfo  [0]{\@secondoftwo}%
\providecommand \bibfield  [0]{\@secondoftwo}%
\providecommand \translation [1]{[#1]}%
\providecommand \BibitemOpen [0]{}%
\providecommand \bibitemStop [0]{}%
\providecommand \bibitemNoStop [0]{.\EOS\space}%
\providecommand \EOS [0]{\spacefactor3000\relax}%
\providecommand \BibitemShut  [1]{\csname bibitem#1\endcsname}%
\let\auto@bib@innerbib\@empty
\bibitem [{\citenamefont {Kapitza}(1938)}]{1938KapitzaNature}%
  \BibitemOpen
  \bibfield  {author} {\bibinfo {author} {\bibfnamefont {P.}~\bibnamefont
  {Kapitza}},\ }\href {\doibase 10.1038/141074a0} {\bibfield  {journal}
  {\bibinfo  {journal} {Nature}\ }\textbf {\bibinfo {volume} {141}},\ \bibinfo
  {pages} {74} (\bibinfo {year} {1938})}\BibitemShut {NoStop}%
\bibitem [{\citenamefont {Allen}\ and\ \citenamefont
  {Misener}(1938)}]{1938AllenNature}%
  \BibitemOpen
  \bibfield  {author} {\bibinfo {author} {\bibfnamefont {J.~F.}\ \bibnamefont
  {Allen}}\ and\ \bibinfo {author} {\bibfnamefont {A.~D.}\ \bibnamefont
  {Misener}},\ }\href {\doibase 10.1038/142643a0} {\bibfield  {journal}
  {\bibinfo  {journal} {Nature}\ }\textbf {\bibinfo {volume} {142}},\ \bibinfo
  {pages} {643} (\bibinfo {year} {1938})}\BibitemShut {NoStop}%
\bibitem [{\citenamefont {Onnes}(1911)}]{1911Onnes}%
  \BibitemOpen
  \bibfield  {author} {\bibinfo {author} {\bibfnamefont {H.~K.}\ \bibnamefont
  {Onnes}},\ }\href@noop {} {\bibfield  {journal} {\bibinfo  {journal} {Commun.
  Phys. Lab. Univ. Leiden}\ }\textbf {\bibinfo {volume} {12}},\ \bibinfo
  {pages} {120} (\bibinfo {year} {1911})}\BibitemShut {NoStop}%
\bibitem [{\citenamefont {Bardeen}\ \emph {et~al.}(1957)\citenamefont
  {Bardeen}, \citenamefont {Cooper},\ and\ \citenamefont
  {Schrieffer}}]{1957BardeenPhysRev}%
  \BibitemOpen
  \bibfield  {author} {\bibinfo {author} {\bibfnamefont {J.}~\bibnamefont
  {Bardeen}}, \bibinfo {author} {\bibfnamefont {L.~N.}\ \bibnamefont {Cooper}},
  \ and\ \bibinfo {author} {\bibfnamefont {J.~R.}\ \bibnamefont {Schrieffer}},\
  }\href {\doibase 10.1103/PhysRev.108.1175} {\bibfield  {journal} {\bibinfo
  {journal} {Phys. Rev.}\ }\textbf {\bibinfo {volume} {108}},\ \bibinfo {pages}
  {1175} (\bibinfo {year} {1957})}\BibitemShut {NoStop}%
\bibitem [{\citenamefont {Stormer}\ \emph {et~al.}(1999)\citenamefont
  {Stormer}, \citenamefont {Tsui},\ and\ \citenamefont
  {Gossard}}]{1999StormerRMP}%
  \BibitemOpen
  \bibfield  {author} {\bibinfo {author} {\bibfnamefont {H.~L.}\ \bibnamefont
  {Stormer}}, \bibinfo {author} {\bibfnamefont {D.~C.}\ \bibnamefont {Tsui}}, \
  and\ \bibinfo {author} {\bibfnamefont {A.~C.}\ \bibnamefont {Gossard}},\
  }\href {\doibase 10.1103/RevModPhys.71.S298} {\bibfield  {journal} {\bibinfo
  {journal} {Rev. Mod. Phys.}\ }\textbf {\bibinfo {volume} {71}},\ \bibinfo
  {pages} {S298} (\bibinfo {year} {1999})}\BibitemShut {NoStop}%
\bibitem [{\citenamefont {Baibich}\ \emph {et~al.}(1988)\citenamefont
  {Baibich}, \citenamefont {Broto}, \citenamefont {Fert}, \citenamefont
  {Van~Dau}, \citenamefont {Petroff}, \citenamefont {Etienne}, \citenamefont
  {Creuzet}, \citenamefont {Friederich},\ and\ \citenamefont
  {Chazelas}}]{1988BaibichPRL}%
  \BibitemOpen
  \bibfield  {author} {\bibinfo {author} {\bibfnamefont {M.~N.}\ \bibnamefont
  {Baibich}}, \bibinfo {author} {\bibfnamefont {J.~M.}\ \bibnamefont {Broto}},
  \bibinfo {author} {\bibfnamefont {A.}~\bibnamefont {Fert}}, \bibinfo {author}
  {\bibfnamefont {F.~N.}\ \bibnamefont {Van~Dau}}, \bibinfo {author}
  {\bibfnamefont {F.}~\bibnamefont {Petroff}}, \bibinfo {author} {\bibfnamefont
  {P.}~\bibnamefont {Etienne}}, \bibinfo {author} {\bibfnamefont
  {G.}~\bibnamefont {Creuzet}}, \bibinfo {author} {\bibfnamefont
  {A.}~\bibnamefont {Friederich}}, \ and\ \bibinfo {author} {\bibfnamefont
  {J.}~\bibnamefont {Chazelas}},\ }\href {\doibase 10.1103/PhysRevLett.61.2472}
  {\bibfield  {journal} {\bibinfo  {journal} {Phys. Rev. Lett.}\ }\textbf
  {\bibinfo {volume} {61}},\ \bibinfo {pages} {2472} (\bibinfo {year}
  {1988})}\BibitemShut {NoStop}%
\bibitem [{\citenamefont {Binasch}\ \emph {et~al.}(1989)\citenamefont
  {Binasch}, \citenamefont {Gr\"unberg}, \citenamefont {Saurenbach},\ and\
  \citenamefont {Zinn}}]{1989BinaschPRB}%
  \BibitemOpen
  \bibfield  {author} {\bibinfo {author} {\bibfnamefont {G.}~\bibnamefont
  {Binasch}}, \bibinfo {author} {\bibfnamefont {P.}~\bibnamefont {Gr\"unberg}},
  \bibinfo {author} {\bibfnamefont {F.}~\bibnamefont {Saurenbach}}, \ and\
  \bibinfo {author} {\bibfnamefont {W.}~\bibnamefont {Zinn}},\ }\href {\doibase
  10.1103/PhysRevB.39.4828} {\bibfield  {journal} {\bibinfo  {journal} {Phys.
  Rev. B}\ }\textbf {\bibinfo {volume} {39}},\ \bibinfo {pages} {4828}
  (\bibinfo {year} {1989})}\BibitemShut {NoStop}%
\bibitem [{\citenamefont {Georges}\ \emph {et~al.}(1996)\citenamefont
  {Georges}, \citenamefont {Kotliar}, \citenamefont {Krauth},\ and\
  \citenamefont {Rozenberg}}]{1996GeorgesRMP}%
  \BibitemOpen
  \bibfield  {author} {\bibinfo {author} {\bibfnamefont {A.}~\bibnamefont
  {Georges}}, \bibinfo {author} {\bibfnamefont {G.}~\bibnamefont {Kotliar}},
  \bibinfo {author} {\bibfnamefont {W.}~\bibnamefont {Krauth}}, \ and\ \bibinfo
  {author} {\bibfnamefont {M.~J.}\ \bibnamefont {Rozenberg}},\ }\href {\doibase
  10.1103/RevModPhys.68.13} {\bibfield  {journal} {\bibinfo  {journal} {Rev.
  Mod. Phys.}\ }\textbf {\bibinfo {volume} {68}},\ \bibinfo {pages} {13}
  (\bibinfo {year} {1996})}\BibitemShut {NoStop}%
\bibitem [{\citenamefont {Dalfovo}\ \emph {et~al.}(1999)\citenamefont
  {Dalfovo}, \citenamefont {Giorgini}, \citenamefont {Pitaevskii},\ and\
  \citenamefont {Stringari}}]{1999DalfovoRMP}%
  \BibitemOpen
  \bibfield  {author} {\bibinfo {author} {\bibfnamefont {F.}~\bibnamefont
  {Dalfovo}}, \bibinfo {author} {\bibfnamefont {S.}~\bibnamefont {Giorgini}},
  \bibinfo {author} {\bibfnamefont {L.~P.}\ \bibnamefont {Pitaevskii}}, \ and\
  \bibinfo {author} {\bibfnamefont {S.}~\bibnamefont {Stringari}},\ }\href
  {\doibase 10.1103/RevModPhys.71.463} {\bibfield  {journal} {\bibinfo
  {journal} {Rev. Mod. Phys.}\ }\textbf {\bibinfo {volume} {71}},\ \bibinfo
  {pages} {463} (\bibinfo {year} {1999})}\BibitemShut {NoStop}%
\bibitem [{\citenamefont {Bender}\ \emph {et~al.}(2003)\citenamefont {Bender},
  \citenamefont {Heenen},\ and\ \citenamefont {Reinhard}}]{2003BenderRMP}%
  \BibitemOpen
  \bibfield  {author} {\bibinfo {author} {\bibfnamefont {M.}~\bibnamefont
  {Bender}}, \bibinfo {author} {\bibfnamefont {P.-H.}\ \bibnamefont {Heenen}},
  \ and\ \bibinfo {author} {\bibfnamefont {P.-G.}\ \bibnamefont {Reinhard}},\
  }\href {\doibase 10.1103/RevModPhys.75.121} {\bibfield  {journal} {\bibinfo
  {journal} {Rev. Mod. Phys.}\ }\textbf {\bibinfo {volume} {75}},\ \bibinfo
  {pages} {121} (\bibinfo {year} {2003})}\BibitemShut {NoStop}%
\bibitem [{\citenamefont {Jones}(2015)}]{2015JonesRMP}%
  \BibitemOpen
  \bibfield  {author} {\bibinfo {author} {\bibfnamefont {R.~O.}\ \bibnamefont
  {Jones}},\ }\href {\doibase 10.1103/RevModPhys.87.897} {\bibfield  {journal}
  {\bibinfo  {journal} {Rev. Mod. Phys.}\ }\textbf {\bibinfo {volume} {87}},\
  \bibinfo {pages} {897} (\bibinfo {year} {2015})}\BibitemShut {NoStop}%
\bibitem [{\citenamefont {Fisher}(1998)}]{1998FisherRMP}%
  \BibitemOpen
  \bibfield  {author} {\bibinfo {author} {\bibfnamefont {M.~E.}\ \bibnamefont
  {Fisher}},\ }\href {\doibase 10.1103/RevModPhys.70.653} {\bibfield  {journal}
  {\bibinfo  {journal} {Rev. Mod. Phys.}\ }\textbf {\bibinfo {volume} {70}},\
  \bibinfo {pages} {653} (\bibinfo {year} {1998})}\BibitemShut {NoStop}%
\bibitem [{\citenamefont {Murayama}(2001)}]{BookMurayama2001}%
  \BibitemOpen
  \bibfield  {author} {\bibinfo {author} {\bibfnamefont {Y.}~\bibnamefont
  {Murayama}},\ }\href@noop {} {\emph {\bibinfo {title} {Mesoscopic Systems:
  Fundamentals and Applications}}}\ (\bibinfo  {publisher} {Wiley-VCH},\
  \bibinfo {address} {Weinheim},\ \bibinfo {year} {2001})\BibitemShut {NoStop}%
\bibitem [{\citenamefont {Cheinet}\ \emph {et~al.}(2008)\citenamefont
  {Cheinet}, \citenamefont {Trotzky}, \citenamefont {Feld}, \citenamefont
  {Schnorrberger}, \citenamefont {Moreno-Cardoner}, \citenamefont {F\"olling},\
  and\ \citenamefont {Bloch}}]{2008CheinetPRL}%
  \BibitemOpen
  \bibfield  {author} {\bibinfo {author} {\bibfnamefont {P.}~\bibnamefont
  {Cheinet}}, \bibinfo {author} {\bibfnamefont {S.}~\bibnamefont {Trotzky}},
  \bibinfo {author} {\bibfnamefont {M.}~\bibnamefont {Feld}}, \bibinfo {author}
  {\bibfnamefont {U.}~\bibnamefont {Schnorrberger}}, \bibinfo {author}
  {\bibfnamefont {M.}~\bibnamefont {Moreno-Cardoner}}, \bibinfo {author}
  {\bibfnamefont {S.}~\bibnamefont {F\"olling}}, \ and\ \bibinfo {author}
  {\bibfnamefont {I.}~\bibnamefont {Bloch}},\ }\href {\doibase
  10.1103/PhysRevLett.101.090404} {\bibfield  {journal} {\bibinfo  {journal}
  {Phys. Rev. Lett.}\ }\textbf {\bibinfo {volume} {101}},\ \bibinfo {pages}
  {090404} (\bibinfo {year} {2008})}\BibitemShut {NoStop}%
\bibitem [{\citenamefont {Serwane}\ \emph {et~al.}(2011)\citenamefont
  {Serwane}, \citenamefont {Zürn}, \citenamefont {Lompe}, \citenamefont
  {Ottenstein}, \citenamefont {Wenz},\ and\ \citenamefont
  {Jochim}}]{2011SerwaneScience}%
  \BibitemOpen
  \bibfield  {author} {\bibinfo {author} {\bibfnamefont {F.}~\bibnamefont
  {Serwane}}, \bibinfo {author} {\bibfnamefont {G.}~\bibnamefont {Zürn}},
  \bibinfo {author} {\bibfnamefont {T.}~\bibnamefont {Lompe}}, \bibinfo
  {author} {\bibfnamefont {T.~B.}\ \bibnamefont {Ottenstein}}, \bibinfo
  {author} {\bibfnamefont {A.~N.}\ \bibnamefont {Wenz}}, \ and\ \bibinfo
  {author} {\bibfnamefont {S.}~\bibnamefont {Jochim}},\ }\href {\doibase
  10.1126/science.1201351} {\bibfield  {journal} {\bibinfo  {journal}
  {Science}\ }\textbf {\bibinfo {volume} {332}},\ \bibinfo {pages} {336}
  (\bibinfo {year} {2011})}\BibitemShut {NoStop}%
\bibitem [{\citenamefont {Wenz}\ \emph {et~al.}(2013)\citenamefont {Wenz},
  \citenamefont {Zürn}, \citenamefont {Murmann}, \citenamefont {Brouzos},
  \citenamefont {Lompe},\ and\ \citenamefont {Jochim}}]{2013WenzScience}%
  \BibitemOpen
  \bibfield  {author} {\bibinfo {author} {\bibfnamefont {A.~N.}\ \bibnamefont
  {Wenz}}, \bibinfo {author} {\bibfnamefont {G.}~\bibnamefont {Zürn}},
  \bibinfo {author} {\bibfnamefont {S.}~\bibnamefont {Murmann}}, \bibinfo
  {author} {\bibfnamefont {I.}~\bibnamefont {Brouzos}}, \bibinfo {author}
  {\bibfnamefont {T.}~\bibnamefont {Lompe}}, \ and\ \bibinfo {author}
  {\bibfnamefont {S.}~\bibnamefont {Jochim}},\ }\href {\doibase
  10.1126/science.1240516} {\bibfield  {journal} {\bibinfo  {journal}
  {Science}\ }\textbf {\bibinfo {volume} {342}},\ \bibinfo {pages} {457}
  (\bibinfo {year} {2013})}\BibitemShut {NoStop}%
\bibitem [{\citenamefont {Blume}(2012)}]{2012BlumeRPP}%
  \BibitemOpen
  \bibfield  {author} {\bibinfo {author} {\bibfnamefont {D.}~\bibnamefont
  {Blume}},\ }\href {\doibase 10.1088/0034-4885/75/4/046401} {\bibfield
  {journal} {\bibinfo  {journal} {Reports on Progress in Physics}\ }\textbf
  {\bibinfo {volume} {75}},\ \bibinfo {pages} {046401} (\bibinfo {year}
  {2012})}\BibitemShut {NoStop}%
\bibitem [{\citenamefont {{Zinner, Nikolaj Thomas}}(2016)}]{2016ZinnerEPJWoC}%
  \BibitemOpen
  \bibfield  {author} {\bibinfo {author} {\bibnamefont {{Zinner, Nikolaj
  Thomas}}},\ }\href {\doibase 10.1051/epjconf/201611301002} {\bibfield
  {journal} {\bibinfo  {journal} {EPJ Web of Conferences}\ }\textbf {\bibinfo
  {volume} {113}},\ \bibinfo {pages} {01002} (\bibinfo {year}
  {2016})}\BibitemShut {NoStop}%
\bibitem [{\citenamefont {Sowiński}\ and\ \citenamefont {Ángel
  García-March}(2019)}]{2019SowinskiRPP}%
  \BibitemOpen
  \bibfield  {author} {\bibinfo {author} {\bibfnamefont {T.}~\bibnamefont
  {Sowiński}}\ and\ \bibinfo {author} {\bibfnamefont {M.}~\bibnamefont {Ángel
  García-March}},\ }\href {\doibase 10.1088/1361-6633/ab3a80} {\bibfield
  {journal} {\bibinfo  {journal} {Reports on Progress in Physics}\ }\textbf
  {\bibinfo {volume} {82}},\ \bibinfo {pages} {104401} (\bibinfo {year}
  {2019})}\BibitemShut {NoStop}%
\bibitem [{\citenamefont {Holten}\ \emph {et~al.}(2022)\citenamefont {Holten},
  \citenamefont {Bayha}, \citenamefont {Subramanian}, \citenamefont
  {Brandstetter}, \citenamefont {Heintze}, \citenamefont {Lunt}, \citenamefont
  {Preiss},\ and\ \citenamefont {Jochim}}]{2022HoltenNature}%
  \BibitemOpen
  \bibfield  {author} {\bibinfo {author} {\bibfnamefont {M.}~\bibnamefont
  {Holten}}, \bibinfo {author} {\bibfnamefont {L.}~\bibnamefont {Bayha}},
  \bibinfo {author} {\bibfnamefont {K.}~\bibnamefont {Subramanian}}, \bibinfo
  {author} {\bibfnamefont {S.}~\bibnamefont {Brandstetter}}, \bibinfo {author}
  {\bibfnamefont {C.}~\bibnamefont {Heintze}}, \bibinfo {author} {\bibfnamefont
  {P.}~\bibnamefont {Lunt}}, \bibinfo {author} {\bibfnamefont {P.~M.}\
  \bibnamefont {Preiss}}, \ and\ \bibinfo {author} {\bibfnamefont
  {S.}~\bibnamefont {Jochim}},\ }\href {\doibase 10.1038/s41586-022-04678-1}
  {\bibfield  {journal} {\bibinfo  {journal} {Nature}\ }\textbf {\bibinfo
  {volume} {606}},\ \bibinfo {pages} {287} (\bibinfo {year}
  {2022})}\BibitemShut {NoStop}%
\bibitem [{\citenamefont {Laird}\ \emph {et~al.}(2024)\citenamefont {Laird},
  \citenamefont {Mulkerin}, \citenamefont {Wang},\ and\ \citenamefont
  {Davis}}]{2024LairdSciPost}%
  \BibitemOpen
  \bibfield  {author} {\bibinfo {author} {\bibfnamefont {E.~K.}\ \bibnamefont
  {Laird}}, \bibinfo {author} {\bibfnamefont {B.~C.}\ \bibnamefont {Mulkerin}},
  \bibinfo {author} {\bibfnamefont {J.}~\bibnamefont {Wang}}, \ and\ \bibinfo
  {author} {\bibfnamefont {M.~J.}\ \bibnamefont {Davis}},\ }\href {\doibase
  10.21468/SciPostPhys.17.6.163} {\bibfield  {journal} {\bibinfo  {journal}
  {SciPost Phys.}\ }\textbf {\bibinfo {volume} {17}},\ \bibinfo {pages} {163}
  (\bibinfo {year} {2024})}\BibitemShut {NoStop}%
\bibitem [{\citenamefont {Gajda}\ \emph {et~al.}(2016)\citenamefont {Gajda},
  \citenamefont {Mostowski}, \citenamefont {Sowiński},\ and\ \citenamefont
  {Załuska-Kotur}}]{2016GajdaEPL}%
  \BibitemOpen
  \bibfield  {author} {\bibinfo {author} {\bibfnamefont {M.}~\bibnamefont
  {Gajda}}, \bibinfo {author} {\bibfnamefont {J.}~\bibnamefont {Mostowski}},
  \bibinfo {author} {\bibfnamefont {T.}~\bibnamefont {Sowiński}}, \ and\
  \bibinfo {author} {\bibfnamefont {M.}~\bibnamefont {Załuska-Kotur}},\ }\href
  {\doibase 10.1209/0295-5075/115/20012} {\bibfield  {journal} {\bibinfo
  {journal} {Europhysics Letters}\ }\textbf {\bibinfo {volume} {115}},\
  \bibinfo {pages} {20012} (\bibinfo {year} {2016})}\BibitemShut {NoStop}%
\bibitem [{\citenamefont {Holten}\ \emph {et~al.}(2021)\citenamefont {Holten},
  \citenamefont {Bayha}, \citenamefont {Subramanian}, \citenamefont {Heintze},
  \citenamefont {Preiss},\ and\ \citenamefont {Jochim}}]{2021HoltenPRL}%
  \BibitemOpen
  \bibfield  {author} {\bibinfo {author} {\bibfnamefont {M.}~\bibnamefont
  {Holten}}, \bibinfo {author} {\bibfnamefont {L.}~\bibnamefont {Bayha}},
  \bibinfo {author} {\bibfnamefont {K.}~\bibnamefont {Subramanian}}, \bibinfo
  {author} {\bibfnamefont {C.}~\bibnamefont {Heintze}}, \bibinfo {author}
  {\bibfnamefont {P.~M.}\ \bibnamefont {Preiss}}, \ and\ \bibinfo {author}
  {\bibfnamefont {S.}~\bibnamefont {Jochim}},\ }\href {\doibase
  10.1103/PhysRevLett.126.020401} {\bibfield  {journal} {\bibinfo  {journal}
  {Phys. Rev. Lett.}\ }\textbf {\bibinfo {volume} {126}},\ \bibinfo {pages}
  {020401} (\bibinfo {year} {2021})}\BibitemShut {NoStop}%
\bibitem [{\citenamefont {Cazalilla}\ and\ \citenamefont
  {Rey}(2014)}]{2014CazalillaRPP}%
  \BibitemOpen
  \bibfield  {author} {\bibinfo {author} {\bibfnamefont {M.~A.}\ \bibnamefont
  {Cazalilla}}\ and\ \bibinfo {author} {\bibfnamefont {A.~M.}\ \bibnamefont
  {Rey}},\ }\href {\doibase 10.1088/0034-4885/77/12/124401} {\bibfield
  {journal} {\bibinfo  {journal} {Reports on Progress in Physics}\ }\textbf
  {\bibinfo {volume} {77}},\ \bibinfo {pages} {124401} (\bibinfo {year}
  {2014})}\BibitemShut {NoStop}%
\bibitem [{\citenamefont {Pagano}\ \emph {et~al.}(2014)\citenamefont {Pagano},
  \citenamefont {Mancini}, \citenamefont {Cappellini}, \citenamefont
  {Lombardi}, \citenamefont {Sch{\"a}fer}, \citenamefont {Hu}, \citenamefont
  {Liu}, \citenamefont {Catani}, \citenamefont {Sias}, \citenamefont
  {Inguscio},\ and\ \citenamefont {Fallani}}]{2014PaganoNatPhys}%
  \BibitemOpen
  \bibfield  {author} {\bibinfo {author} {\bibfnamefont {G.}~\bibnamefont
  {Pagano}}, \bibinfo {author} {\bibfnamefont {M.}~\bibnamefont {Mancini}},
  \bibinfo {author} {\bibfnamefont {G.}~\bibnamefont {Cappellini}}, \bibinfo
  {author} {\bibfnamefont {P.}~\bibnamefont {Lombardi}}, \bibinfo {author}
  {\bibfnamefont {F.}~\bibnamefont {Sch{\"a}fer}}, \bibinfo {author}
  {\bibfnamefont {H.}~\bibnamefont {Hu}}, \bibinfo {author} {\bibfnamefont
  {X.-J.}\ \bibnamefont {Liu}}, \bibinfo {author} {\bibfnamefont
  {J.}~\bibnamefont {Catani}}, \bibinfo {author} {\bibfnamefont
  {C.}~\bibnamefont {Sias}}, \bibinfo {author} {\bibfnamefont {M.}~\bibnamefont
  {Inguscio}}, \ and\ \bibinfo {author} {\bibfnamefont {L.}~\bibnamefont
  {Fallani}},\ }\href {https://doi.org/10.1038/nphys2878} {\bibfield  {journal}
  {\bibinfo  {journal} {Nature Physics}\ }\textbf {\bibinfo {volume} {10}},\
  \bibinfo {pages} {198 EP } (\bibinfo {year} {2014})}\BibitemShut {NoStop}%
\bibitem [{\citenamefont {Sonderhouse}\ \emph {et~al.}(2020)\citenamefont
  {Sonderhouse}, \citenamefont {Sanner}, \citenamefont {Hutson}, \citenamefont
  {Goban}, \citenamefont {Bilitewski}, \citenamefont {Yan}, \citenamefont
  {Milner}, \citenamefont {Rey},\ and\ \citenamefont
  {Ye}}]{2020SonderhouseNaturePhys}%
  \BibitemOpen
  \bibfield  {author} {\bibinfo {author} {\bibfnamefont {L.}~\bibnamefont
  {Sonderhouse}}, \bibinfo {author} {\bibfnamefont {C.}~\bibnamefont {Sanner}},
  \bibinfo {author} {\bibfnamefont {R.~B.}\ \bibnamefont {Hutson}}, \bibinfo
  {author} {\bibfnamefont {A.}~\bibnamefont {Goban}}, \bibinfo {author}
  {\bibfnamefont {T.}~\bibnamefont {Bilitewski}}, \bibinfo {author}
  {\bibfnamefont {L.}~\bibnamefont {Yan}}, \bibinfo {author} {\bibfnamefont
  {W.~R.}\ \bibnamefont {Milner}}, \bibinfo {author} {\bibfnamefont {A.~M.}\
  \bibnamefont {Rey}}, \ and\ \bibinfo {author} {\bibfnamefont
  {J.}~\bibnamefont {Ye}},\ }\href {\doibase 10.1038/s41567-020-0986-6}
  {\bibfield  {journal} {\bibinfo  {journal} {Nature Physics}\ }\textbf
  {\bibinfo {volume} {16}},\ \bibinfo {pages} {1216} (\bibinfo {year}
  {2020})}\BibitemShut {NoStop}%
\bibitem [{\citenamefont {Liu}\ \emph {et~al.}(2008)\citenamefont {Liu},
  \citenamefont {Hu},\ and\ \citenamefont {Drummond}}]{2008LiaPRA}%
  \BibitemOpen
  \bibfield  {author} {\bibinfo {author} {\bibfnamefont {X.-J.}\ \bibnamefont
  {Liu}}, \bibinfo {author} {\bibfnamefont {H.}~\bibnamefont {Hu}}, \ and\
  \bibinfo {author} {\bibfnamefont {P.~D.}\ \bibnamefont {Drummond}},\ }\href
  {\doibase 10.1103/PhysRevA.77.013622} {\bibfield  {journal} {\bibinfo
  {journal} {Phys. Rev. A}\ }\textbf {\bibinfo {volume} {77}},\ \bibinfo
  {pages} {013622} (\bibinfo {year} {2008})}\BibitemShut {NoStop}%
\bibitem [{\citenamefont {Jiang}\ \emph {et~al.}(2009)\citenamefont {Jiang},
  \citenamefont {Cao},\ and\ \citenamefont {Wang}}]{2009JiangEPL}%
  \BibitemOpen
  \bibfield  {author} {\bibinfo {author} {\bibfnamefont {Y.}~\bibnamefont
  {Jiang}}, \bibinfo {author} {\bibfnamefont {J.}~\bibnamefont {Cao}}, \ and\
  \bibinfo {author} {\bibfnamefont {Y.}~\bibnamefont {Wang}},\ }\href {\doibase
  10.1209/0295-5075/87/10006} {\bibfield  {journal} {\bibinfo  {journal}
  {Europhysics Letters}\ }\textbf {\bibinfo {volume} {87}},\ \bibinfo {pages}
  {10006} (\bibinfo {year} {2009})}\BibitemShut {NoStop}%
\bibitem [{\citenamefont {Guan}\ \emph {et~al.}(2010)\citenamefont {Guan},
  \citenamefont {Lee}, \citenamefont {Batchelor}, \citenamefont {Yin},\ and\
  \citenamefont {Chen}}]{2010GuanPRA}%
  \BibitemOpen
  \bibfield  {author} {\bibinfo {author} {\bibfnamefont {X.~W.}\ \bibnamefont
  {Guan}}, \bibinfo {author} {\bibfnamefont {J.-Y.}\ \bibnamefont {Lee}},
  \bibinfo {author} {\bibfnamefont {M.~T.}\ \bibnamefont {Batchelor}}, \bibinfo
  {author} {\bibfnamefont {X.-G.}\ \bibnamefont {Yin}}, \ and\ \bibinfo
  {author} {\bibfnamefont {S.}~\bibnamefont {Chen}},\ }\href {\doibase
  10.1103/PhysRevA.82.021606} {\bibfield  {journal} {\bibinfo  {journal} {Phys.
  Rev. A}\ }\textbf {\bibinfo {volume} {82}},\ \bibinfo {pages} {021606}
  (\bibinfo {year} {2010})}\BibitemShut {NoStop}%
\bibitem [{\citenamefont {Lee}\ \emph {et~al.}(2011)\citenamefont {Lee},
  \citenamefont {Guan},\ and\ \citenamefont {Batchelor}}]{2011LeeJPA}%
  \BibitemOpen
  \bibfield  {author} {\bibinfo {author} {\bibfnamefont {J.~Y.}\ \bibnamefont
  {Lee}}, \bibinfo {author} {\bibfnamefont {X.~W.}\ \bibnamefont {Guan}}, \
  and\ \bibinfo {author} {\bibfnamefont {M.~T.}\ \bibnamefont {Batchelor}},\
  }\href {\doibase 10.1088/1751-8113/44/16/165002} {\bibfield  {journal}
  {\bibinfo  {journal} {Journal of Physics A: Mathematical and Theoretical}\
  }\textbf {\bibinfo {volume} {44}},\ \bibinfo {pages} {165002} (\bibinfo
  {year} {2011})}\BibitemShut {NoStop}%
\bibitem [{\citenamefont {Kuhn}\ and\ \citenamefont
  {Foerster}(2012)}]{2012KuhnNJP}%
  \BibitemOpen
  \bibfield  {author} {\bibinfo {author} {\bibfnamefont {C.~C.~N.}\
  \bibnamefont {Kuhn}}\ and\ \bibinfo {author} {\bibfnamefont {A.}~\bibnamefont
  {Foerster}},\ }\href {\doibase 10.1088/1367-2630/14/1/013008} {\bibfield
  {journal} {\bibinfo  {journal} {New Journal of Physics}\ }\textbf {\bibinfo
  {volume} {14}},\ \bibinfo {pages} {013008} (\bibinfo {year}
  {2012})}\BibitemShut {NoStop}%
\bibitem [{\citenamefont {Guan}\ \emph {et~al.}(2012)\citenamefont {Guan},
  \citenamefont {Ma},\ and\ \citenamefont {Wilson}}]{2012GuanPRA}%
  \BibitemOpen
  \bibfield  {author} {\bibinfo {author} {\bibfnamefont {X.-W.}\ \bibnamefont
  {Guan}}, \bibinfo {author} {\bibfnamefont {Z.-Q.}\ \bibnamefont {Ma}}, \ and\
  \bibinfo {author} {\bibfnamefont {B.}~\bibnamefont {Wilson}},\ }\href
  {\doibase 10.1103/PhysRevA.85.033633} {\bibfield  {journal} {\bibinfo
  {journal} {Phys. Rev. A}\ }\textbf {\bibinfo {volume} {85}},\ \bibinfo
  {pages} {033633} (\bibinfo {year} {2012})}\BibitemShut {NoStop}%
\bibitem [{\citenamefont {Jiang}\ \emph {et~al.}(2016)\citenamefont {Jiang},
  \citenamefont {He},\ and\ \citenamefont {Guan}}]{2016JiangJPA}%
  \BibitemOpen
  \bibfield  {author} {\bibinfo {author} {\bibfnamefont {Y.}~\bibnamefont
  {Jiang}}, \bibinfo {author} {\bibfnamefont {P.}~\bibnamefont {He}}, \ and\
  \bibinfo {author} {\bibfnamefont {X.-W.}\ \bibnamefont {Guan}},\ }\href
  {\doibase 10.1088/1751-8113/49/17/174005} {\bibfield  {journal} {\bibinfo
  {journal} {Journal of Physics A: Mathematical and Theoretical}\ }\textbf
  {\bibinfo {volume} {49}},\ \bibinfo {pages} {174005} (\bibinfo {year}
  {2016})}\BibitemShut {NoStop}%
\bibitem [{\citenamefont {Laird}\ \emph {et~al.}(2017)\citenamefont {Laird},
  \citenamefont {Shi}, \citenamefont {Parish},\ and\ \citenamefont
  {Levinsen}}]{2017LairdPRA}%
  \BibitemOpen
  \bibfield  {author} {\bibinfo {author} {\bibfnamefont {E.~K.}\ \bibnamefont
  {Laird}}, \bibinfo {author} {\bibfnamefont {Z.-Y.}\ \bibnamefont {Shi}},
  \bibinfo {author} {\bibfnamefont {M.~M.}\ \bibnamefont {Parish}}, \ and\
  \bibinfo {author} {\bibfnamefont {J.}~\bibnamefont {Levinsen}},\ }\href
  {\doibase 10.1103/PhysRevA.96.032701} {\bibfield  {journal} {\bibinfo
  {journal} {Phys. Rev. A}\ }\textbf {\bibinfo {volume} {96}},\ \bibinfo
  {pages} {032701} (\bibinfo {year} {2017})}\BibitemShut {NoStop}%
\bibitem [{\citenamefont {Dobrzyniecki}\ and\ \citenamefont
  {Sowiński}(2020)}]{2020DobrzynieckiAdvQT}%
  \BibitemOpen
  \bibfield  {author} {\bibinfo {author} {\bibfnamefont {J.}~\bibnamefont
  {Dobrzyniecki}}\ and\ \bibinfo {author} {\bibfnamefont {T.}~\bibnamefont
  {Sowiński}},\ }\href {\doibase https://doi.org/10.1002/qute.202000010}
  {\bibfield  {journal} {\bibinfo  {journal} {Advanced Quantum Technologies}\
  }\textbf {\bibinfo {volume} {3}},\ \bibinfo {pages} {2000010} (\bibinfo
  {year} {2020})}\BibitemShut {NoStop}%
\bibitem [{\citenamefont {Anh-Tai}\ \emph {et~al.}(2024)\citenamefont
  {Anh-Tai}, \citenamefont {Fogarty}, \citenamefont
  {de~Mar\'{\i}a-Garc\'{\i}a}, \citenamefont {Busch},\ and\ \citenamefont
  {Garc\'{\i}a-March}}]{2024TaiPRR}%
  \BibitemOpen
  \bibfield  {author} {\bibinfo {author} {\bibfnamefont {T.~D.}\ \bibnamefont
  {Anh-Tai}}, \bibinfo {author} {\bibfnamefont {T.}~\bibnamefont {Fogarty}},
  \bibinfo {author} {\bibfnamefont {S.}~\bibnamefont
  {de~Mar\'{\i}a-Garc\'{\i}a}}, \bibinfo {author} {\bibfnamefont
  {T.}~\bibnamefont {Busch}}, \ and\ \bibinfo {author} {\bibfnamefont {M.~A.}\
  \bibnamefont {Garc\'{\i}a-March}},\ }\href {\doibase
  10.1103/PhysRevResearch.6.043042} {\bibfield  {journal} {\bibinfo  {journal}
  {Phys. Rev. Res.}\ }\textbf {\bibinfo {volume} {6}},\ \bibinfo {pages}
  {043042} (\bibinfo {year} {2024})}\BibitemShut {NoStop}%
\bibitem [{\citenamefont {Silva-Valencia}\ and\ \citenamefont
  {Mendoza-Arenas}(2025)}]{2025SilvaPRA}%
  \BibitemOpen
  \bibfield  {author} {\bibinfo {author} {\bibfnamefont {J.}~\bibnamefont
  {Silva-Valencia}}\ and\ \bibinfo {author} {\bibfnamefont {J.~J.}\
  \bibnamefont {Mendoza-Arenas}},\ }\href {\doibase
  10.1103/PhysRevA.111.053316} {\bibfield  {journal} {\bibinfo  {journal}
  {Phys. Rev. A}\ }\textbf {\bibinfo {volume} {111}},\ \bibinfo {pages}
  {053316} (\bibinfo {year} {2025})}\BibitemShut {NoStop}%
\bibitem [{\citenamefont {Tran}\ \emph {et~al.}(2025)\citenamefont {Tran},
  \citenamefont {Garcia~March}, \citenamefont {Busch},\ and\ \citenamefont
  {Fogarty}}]{2025TranNJP}%
  \BibitemOpen
  \bibfield  {author} {\bibinfo {author} {\bibfnamefont {D.~A.-T.}\
  \bibnamefont {Tran}}, \bibinfo {author} {\bibfnamefont {M.~A.}\ \bibnamefont
  {Garcia~March}}, \bibinfo {author} {\bibfnamefont {T.}~\bibnamefont {Busch}},
  \ and\ \bibinfo {author} {\bibfnamefont {T.}~\bibnamefont {Fogarty}},\ }\href
  {http://iopscience.iop.org/article/10.1088/1367-2630/ade89e} {\bibfield
  {journal} {\bibinfo  {journal} {New Journal of Physics}\ } (\bibinfo {year}
  {2025})}\BibitemShut {NoStop}%
\bibitem [{\citenamefont {Ottenstein}\ \emph {et~al.}(2008)\citenamefont
  {Ottenstein}, \citenamefont {Lompe}, \citenamefont {Kohnen}, \citenamefont
  {Wenz},\ and\ \citenamefont {Jochim}}]{2008OttensteinPRL}%
  \BibitemOpen
  \bibfield  {author} {\bibinfo {author} {\bibfnamefont {T.~B.}\ \bibnamefont
  {Ottenstein}}, \bibinfo {author} {\bibfnamefont {T.}~\bibnamefont {Lompe}},
  \bibinfo {author} {\bibfnamefont {M.}~\bibnamefont {Kohnen}}, \bibinfo
  {author} {\bibfnamefont {A.~N.}\ \bibnamefont {Wenz}}, \ and\ \bibinfo
  {author} {\bibfnamefont {S.}~\bibnamefont {Jochim}},\ }\href {\doibase
  10.1103/PhysRevLett.101.203202} {\bibfield  {journal} {\bibinfo  {journal}
  {Phys. Rev. Lett.}\ }\textbf {\bibinfo {volume} {101}},\ \bibinfo {pages}
  {203202} (\bibinfo {year} {2008})}\BibitemShut {NoStop}%
\bibitem [{\citenamefont {Schumacher}\ \emph {et~al.}(2023)\citenamefont
  {Schumacher}, \citenamefont {Mäkinen}, \citenamefont {Ji}, \citenamefont
  {Assumpção}, \citenamefont {Chen}, \citenamefont {Huang}, \citenamefont
  {Vivanco},\ and\ \citenamefont {Navon}}]{2023SchumacherARXiV}%
  \BibitemOpen
  \bibfield  {author} {\bibinfo {author} {\bibfnamefont {G.~L.}\ \bibnamefont
  {Schumacher}}, \bibinfo {author} {\bibfnamefont {J.~T.}\ \bibnamefont
  {Mäkinen}}, \bibinfo {author} {\bibfnamefont {Y.}~\bibnamefont {Ji}},
  \bibinfo {author} {\bibfnamefont {G.~G.~T.}\ \bibnamefont {Assumpção}},
  \bibinfo {author} {\bibfnamefont {J.}~\bibnamefont {Chen}}, \bibinfo {author}
  {\bibfnamefont {S.}~\bibnamefont {Huang}}, \bibinfo {author} {\bibfnamefont
  {F.~J.}\ \bibnamefont {Vivanco}}, \ and\ \bibinfo {author} {\bibfnamefont
  {N.}~\bibnamefont {Navon}},\ }\href {https://arxiv.org/abs/2301.02237}
  {\enquote {\bibinfo {title} {Observation of anomalous decay of a polarized
  three-component fermi gas},}\ } (\bibinfo {year} {2023}),\ \Eprint
  {http://arxiv.org/abs/2301.02237} {arXiv:2301.02237 [cond-mat.quant-gas]}
  \BibitemShut {NoStop}%
\bibitem [{\citenamefont {W\l{}odzy\ifmmode~\acute{n}\else
  \'{n}\fi{}ski}(2022)}]{2022WlodzynskiPRA}%
  \BibitemOpen
  \bibfield  {author} {\bibinfo {author} {\bibfnamefont {D.}~\bibnamefont
  {W\l{}odzy\ifmmode~\acute{n}\else \'{n}\fi{}ski}},\ }\href {\doibase
  10.1103/PhysRevA.106.033306} {\bibfield  {journal} {\bibinfo  {journal}
  {Phys. Rev. A}\ }\textbf {\bibinfo {volume} {106}},\ \bibinfo {pages}
  {033306} (\bibinfo {year} {2022})}\BibitemShut {NoStop}%
\bibitem [{\citenamefont {Chin}\ \emph {et~al.}(2010)\citenamefont {Chin},
  \citenamefont {Grimm}, \citenamefont {Julienne},\ and\ \citenamefont
  {Tiesinga}}]{2010ChinRMP}%
  \BibitemOpen
  \bibfield  {author} {\bibinfo {author} {\bibfnamefont {C.}~\bibnamefont
  {Chin}}, \bibinfo {author} {\bibfnamefont {R.}~\bibnamefont {Grimm}},
  \bibinfo {author} {\bibfnamefont {P.}~\bibnamefont {Julienne}}, \ and\
  \bibinfo {author} {\bibfnamefont {E.}~\bibnamefont {Tiesinga}},\ }\href
  {\doibase 10.1103/RevModPhys.82.1225} {\bibfield  {journal} {\bibinfo
  {journal} {Rev. Mod. Phys.}\ }\textbf {\bibinfo {volume} {82}},\ \bibinfo
  {pages} {1225} (\bibinfo {year} {2010})}\BibitemShut {NoStop}%
\bibitem [{\citenamefont {Olshanii}(1998)}]{1998OlshaniiPRL}%
  \BibitemOpen
  \bibfield  {author} {\bibinfo {author} {\bibfnamefont {M.}~\bibnamefont
  {Olshanii}},\ }\href {\doibase 10.1103/PhysRevLett.81.938} {\bibfield
  {journal} {\bibinfo  {journal} {Phys. Rev. Lett.}\ }\textbf {\bibinfo
  {volume} {81}},\ \bibinfo {pages} {938} (\bibinfo {year} {1998})}\BibitemShut
  {NoStop}%
\bibitem [{\citenamefont {Xu}\ \emph {et~al.}(2017)\citenamefont {Xu},
  \citenamefont {Feng},\ and\ \citenamefont {Gu}}]{2017XuAnnPhys}%
  \BibitemOpen
  \bibfield  {author} {\bibinfo {author} {\bibfnamefont {J.}~\bibnamefont
  {Xu}}, \bibinfo {author} {\bibfnamefont {T.}~\bibnamefont {Feng}}, \ and\
  \bibinfo {author} {\bibfnamefont {Q.}~\bibnamefont {Gu}},\ }\href {\doibase
  https://doi.org/10.1016/j.aop.2017.02.003} {\bibfield  {journal} {\bibinfo
  {journal} {Annals of Physics}\ }\textbf {\bibinfo {volume} {379}},\ \bibinfo
  {pages} {175} (\bibinfo {year} {2017})}\BibitemShut {NoStop}%
\bibitem [{\citenamefont {Haugset}\ and\ \citenamefont
  {Haugerud}(1998)}]{1998HaugsetPRA}%
  \BibitemOpen
  \bibfield  {author} {\bibinfo {author} {\bibfnamefont {T.}~\bibnamefont
  {Haugset}}\ and\ \bibinfo {author} {\bibfnamefont {H.}~\bibnamefont
  {Haugerud}},\ }\href {\doibase 10.1103/PhysRevA.57.3809} {\bibfield
  {journal} {\bibinfo  {journal} {Phys. Rev. A}\ }\textbf {\bibinfo {volume}
  {57}},\ \bibinfo {pages} {3809} (\bibinfo {year} {1998})}\BibitemShut
  {NoStop}%
\bibitem [{\citenamefont {Płodzień}\ \emph {et~al.}(2018)\citenamefont
  {Płodzień}, \citenamefont {Wiater}, \citenamefont {Chrostowski},\ and\
  \citenamefont {Sowiński}}]{2018PlodzienARXIV}%
  \BibitemOpen
  \bibfield  {author} {\bibinfo {author} {\bibfnamefont {M.}~\bibnamefont
  {Płodzień}}, \bibinfo {author} {\bibfnamefont {D.}~\bibnamefont {Wiater}},
  \bibinfo {author} {\bibfnamefont {A.}~\bibnamefont {Chrostowski}}, \ and\
  \bibinfo {author} {\bibfnamefont {T.}~\bibnamefont {Sowiński}},\ }\href
  {https://arxiv.org/abs/1803.08387} {} (\bibinfo {year} {2018}),\ \Eprint
  {http://arxiv.org/abs/1803.08387} {arXiv:1803.08387 [cond-mat.quant-gas]}
  \BibitemShut {NoStop}%
\bibitem [{\citenamefont {Chrostowski}\ and\ \citenamefont
  {Sowiński}(2019)}]{2019ChrostowskiAPPA}%
  \BibitemOpen
  \bibfield  {author} {\bibinfo {author} {\bibfnamefont {A.}~\bibnamefont
  {Chrostowski}}\ and\ \bibinfo {author} {\bibfnamefont {T.}~\bibnamefont
  {Sowiński}},\ }\href {\doibase 10.12693/aphyspola.136.566} {\bibfield
  {journal} {\bibinfo  {journal} {Acta Physica Polonica A}\ }\textbf {\bibinfo
  {volume} {136}},\ \bibinfo {pages} {566–570} (\bibinfo {year}
  {2019})}\BibitemShut {NoStop}%
\bibitem [{\citenamefont {Rojo-Franc\`as}\ \emph {et~al.}(2022)\citenamefont
  {Rojo-Franc\`as}, \citenamefont {Isaule},\ and\ \citenamefont
  {Juli\'a-D\'{\i}az}}]{2022RojoPRA}%
  \BibitemOpen
  \bibfield  {author} {\bibinfo {author} {\bibfnamefont {A.}~\bibnamefont
  {Rojo-Franc\`as}}, \bibinfo {author} {\bibfnamefont {F.}~\bibnamefont
  {Isaule}}, \ and\ \bibinfo {author} {\bibfnamefont {B.}~\bibnamefont
  {Juli\'a-D\'{\i}az}},\ }\href {\doibase 10.1103/PhysRevA.105.063326}
  {\bibfield  {journal} {\bibinfo  {journal} {Phys. Rev. A}\ }\textbf {\bibinfo
  {volume} {105}},\ \bibinfo {pages} {063326} (\bibinfo {year}
  {2022})}\BibitemShut {NoStop}%
\bibitem [{\citenamefont {Lehoucq}\ \emph {et~al.}(1998)\citenamefont
  {Lehoucq}, \citenamefont {Sorensen},\ and\ \citenamefont
  {Yang}}]{1998BookARPACK}%
  \BibitemOpen
  \bibfield  {author} {\bibinfo {author} {\bibfnamefont {R.~B.}\ \bibnamefont
  {Lehoucq}}, \bibinfo {author} {\bibfnamefont {D.~C.}\ \bibnamefont
  {Sorensen}}, \ and\ \bibinfo {author} {\bibfnamefont {C.}~\bibnamefont
  {Yang}},\ }\href {\doibase 10.1137/1.9780898719628} {\emph {\bibinfo {title}
  {ARPACK Users' Guide}}}\ (\bibinfo  {publisher} {Society for Industrial and
  Applied Mathematics},\ \bibinfo {year} {1998})\BibitemShut {NoStop}%
\bibitem [{\citenamefont {{Volosniev, A.G.}}\ \emph {et~al.}(2015)\citenamefont
  {{Volosniev, A.G.}}, \citenamefont {{Fedorov, D.V.}}, \citenamefont {{Jensen,
  A.S.}},\ and\ \citenamefont {{Zinner, N.T.}}}]{2015VolosnievEPJST}%
  \BibitemOpen
  \bibfield  {author} {\bibinfo {author} {\bibnamefont {{Volosniev, A.G.}}},
  \bibinfo {author} {\bibnamefont {{Fedorov, D.V.}}}, \bibinfo {author}
  {\bibnamefont {{Jensen, A.S.}}}, \ and\ \bibinfo {author} {\bibnamefont
  {{Zinner, N.T.}}},\ }\href {\doibase 10.1140/epjst/e2015-02390-2} {\bibfield
  {journal} {\bibinfo  {journal} {Eur. Phys. J. Special Topics}\ }\textbf
  {\bibinfo {volume} {224}},\ \bibinfo {pages} {585} (\bibinfo {year}
  {2015})}\BibitemShut {NoStop}%
\bibitem [{\citenamefont {{Bellotti, Filipe F.}}\ \emph
  {et~al.}(2017)\citenamefont {{Bellotti, Filipe F.}}, \citenamefont
  {{Dehkharghani, Amin S.}},\ and\ \citenamefont {{Zinner, Nikolaj
  T.}}}]{2017BellottiEPJD}%
  \BibitemOpen
  \bibfield  {author} {\bibinfo {author} {\bibnamefont {{Bellotti, Filipe
  F.}}}, \bibinfo {author} {\bibnamefont {{Dehkharghani, Amin S.}}}, \ and\
  \bibinfo {author} {\bibnamefont {{Zinner, Nikolaj T.}}},\ }\href {\doibase
  10.1140/epjd/e2017-70650-8} {\bibfield  {journal} {\bibinfo  {journal} {Eur.
  Phys. J. D}\ }\textbf {\bibinfo {volume} {71}},\ \bibinfo {pages} {37}
  (\bibinfo {year} {2017})}\BibitemShut {NoStop}%
\bibitem [{\citenamefont {Kościk}(2018)}]{2018KoscikPLA}%
  \BibitemOpen
  \bibfield  {author} {\bibinfo {author} {\bibfnamefont {P.}~\bibnamefont
  {Kościk}},\ }\href {\doibase https://doi.org/10.1016/j.physleta.2018.06.025}
  {\bibfield  {journal} {\bibinfo  {journal} {Physics Letters A}\ }\textbf
  {\bibinfo {volume} {382}},\ \bibinfo {pages} {2561} (\bibinfo {year}
  {2018})}\BibitemShut {NoStop}%
\bibitem [{\citenamefont {Ko{\'s}cik}(2020)}]{2020KoscikFBS}%
  \BibitemOpen
  \bibfield  {author} {\bibinfo {author} {\bibfnamefont {P.}~\bibnamefont
  {Ko{\'s}cik}},\ }\href {\doibase 10.1007/s00601-020-01547-3} {\bibfield
  {journal} {\bibinfo  {journal} {Few-Body Systems}\ }\textbf {\bibinfo
  {volume} {61}},\ \bibinfo {pages} {13} (\bibinfo {year} {2020})}\BibitemShut
  {NoStop}%
\bibitem [{\citenamefont {Teske}\ and\ \citenamefont
  {Sowi\'nski}(2025)}]{zenodo}%
  \BibitemOpen
  \bibfield  {author} {\bibinfo {author} {\bibfnamefont {M.}~\bibnamefont
  {Teske}}\ and\ \bibinfo {author} {\bibfnamefont {T.}~\bibnamefont
  {Sowi\'nski}},\ }\href {\doibase 10.5281/zenodo.15525598} {\enquote {\bibinfo
  {title} {{Supplemental Material for "Impurity immersed in two-component
  few-fermion mixture in one-dimensional harmonic trap"}},}\ }\bibinfo
  {howpublished} {{Zenodo}} (\bibinfo {year} {2025})\BibitemShut {NoStop}%
\bibitem [{\citenamefont {Lindgren}\ \emph {et~al.}(2014)\citenamefont
  {Lindgren}, \citenamefont {Rotureau}, \citenamefont {Forssén}, \citenamefont
  {Volosniev},\ and\ \citenamefont {Zinner}}]{2014LindgrenNJP}%
  \BibitemOpen
  \bibfield  {author} {\bibinfo {author} {\bibfnamefont {E.~J.}\ \bibnamefont
  {Lindgren}}, \bibinfo {author} {\bibfnamefont {J.}~\bibnamefont {Rotureau}},
  \bibinfo {author} {\bibfnamefont {C.}~\bibnamefont {Forssén}}, \bibinfo
  {author} {\bibfnamefont {A.~G.}\ \bibnamefont {Volosniev}}, \ and\ \bibinfo
  {author} {\bibfnamefont {N.~T.}\ \bibnamefont {Zinner}},\ }\href {\doibase
  10.1088/1367-2630/16/6/063003} {\bibfield  {journal} {\bibinfo  {journal}
  {New Journal of Physics}\ }\textbf {\bibinfo {volume} {16}},\ \bibinfo
  {pages} {063003} (\bibinfo {year} {2014})}\BibitemShut {NoStop}%
\bibitem [{\citenamefont {P{\c e}cak}\ \emph {et~al.}(2016)\citenamefont {P{\c
  e}cak}, \citenamefont {Gajda},\ and\ \citenamefont
  {Sowi\'nski}}]{2016PecakNJP}%
  \BibitemOpen
  \bibfield  {author} {\bibinfo {author} {\bibfnamefont {D.}~\bibnamefont {P{\c
  e}cak}}, \bibinfo {author} {\bibfnamefont {M.}~\bibnamefont {Gajda}}, \ and\
  \bibinfo {author} {\bibfnamefont {T.}~\bibnamefont {Sowi\'nski}},\ }\href
  {\doibase 10.1088/1367-2630/18/1/013030} {\bibfield  {journal} {\bibinfo
  {journal} {New Journal of Physics}\ }\textbf {\bibinfo {volume} {18}},\
  \bibinfo {pages} {013030} (\bibinfo {year} {2016})}\BibitemShut {NoStop}%
\bibitem [{\citenamefont {P\ifmmode~\mbox{\k{e}}\else \k{e}\fi{}cak}\ and\
  \citenamefont {Sowi\ifmmode~\acute{n}\else
  \'{n}\fi{}ski}(2016)}]{2016PecakPRA}%
  \BibitemOpen
  \bibfield  {author} {\bibinfo {author} {\bibfnamefont {D.}~\bibnamefont
  {P\ifmmode~\mbox{\k{e}}\else \k{e}\fi{}cak}}\ and\ \bibinfo {author}
  {\bibfnamefont {T.}~\bibnamefont {Sowi\ifmmode~\acute{n}\else
  \'{n}\fi{}ski}},\ }\href {\doibase 10.1103/PhysRevA.94.042118} {\bibfield
  {journal} {\bibinfo  {journal} {Phys. Rev. A}\ }\textbf {\bibinfo {volume}
  {94}},\ \bibinfo {pages} {042118} (\bibinfo {year} {2016})}\BibitemShut
  {NoStop}%
\bibitem [{\citenamefont {You}\ \emph {et~al.}(2007)\citenamefont {You},
  \citenamefont {Li},\ and\ \citenamefont {Gu}}]{2007YouPRE}%
  \BibitemOpen
  \bibfield  {author} {\bibinfo {author} {\bibfnamefont {W.-L.}\ \bibnamefont
  {You}}, \bibinfo {author} {\bibfnamefont {Y.-W.}\ \bibnamefont {Li}}, \ and\
  \bibinfo {author} {\bibfnamefont {S.-J.}\ \bibnamefont {Gu}},\ }\href
  {\doibase 10.1103/PhysRevE.76.022101} {\bibfield  {journal} {\bibinfo
  {journal} {Phys. Rev. E}\ }\textbf {\bibinfo {volume} {76}},\ \bibinfo
  {pages} {022101} (\bibinfo {year} {2007})}\BibitemShut {NoStop}%
\bibitem [{\citenamefont {Wang}\ \emph {et~al.}(2015)\citenamefont {Wang},
  \citenamefont {Liu}, \citenamefont {Imri\ifmmode~\check{s}\else
  \v{s}\fi{}ka}, \citenamefont {Ma},\ and\ \citenamefont
  {Troyer}}]{2015WangPRX}%
  \BibitemOpen
  \bibfield  {author} {\bibinfo {author} {\bibfnamefont {L.}~\bibnamefont
  {Wang}}, \bibinfo {author} {\bibfnamefont {Y.-H.}\ \bibnamefont {Liu}},
  \bibinfo {author} {\bibfnamefont {J.}~\bibnamefont
  {Imri\ifmmode~\check{s}\else \v{s}\fi{}ka}}, \bibinfo {author} {\bibfnamefont
  {P.~N.}\ \bibnamefont {Ma}}, \ and\ \bibinfo {author} {\bibfnamefont
  {M.}~\bibnamefont {Troyer}},\ }\href {\doibase 10.1103/PhysRevX.5.031007}
  {\bibfield  {journal} {\bibinfo  {journal} {Phys. Rev. X}\ }\textbf {\bibinfo
  {volume} {5}},\ \bibinfo {pages} {031007} (\bibinfo {year}
  {2015})}\BibitemShut {NoStop}%
\bibitem [{\citenamefont {Mistakidis}\ \emph {et~al.}(2019)\citenamefont
  {Mistakidis}, \citenamefont {Katsimiga}, \citenamefont {Koutentakis},\ and\
  \citenamefont {Schmelcher}}]{2019MistakidisNJP}%
  \BibitemOpen
  \bibfield  {author} {\bibinfo {author} {\bibfnamefont {S.~I.}\ \bibnamefont
  {Mistakidis}}, \bibinfo {author} {\bibfnamefont {G.~C.}\ \bibnamefont
  {Katsimiga}}, \bibinfo {author} {\bibfnamefont {G.~M.}\ \bibnamefont
  {Koutentakis}}, \ and\ \bibinfo {author} {\bibfnamefont {P.}~\bibnamefont
  {Schmelcher}},\ }\href {\doibase 10.1088/1367-2630/ab1045} {\bibfield
  {journal} {\bibinfo  {journal} {New Journal of Physics}\ }\textbf {\bibinfo
  {volume} {21}},\ \bibinfo {pages} {043032} (\bibinfo {year}
  {2019})}\BibitemShut {NoStop}%
\end{thebibliography}%

\end{document}